\documentclass{brspublic}
\usepackage{epsf,epsfig}
\newcount\timecount
\newcount\hours \newcount\minutes  \newcount\temp \newcount\pmhours
\hours = \time
\divide\hours by 60
\temp = \hours
\multiply\temp by 60
\minutes = \time
\advance\minutes by -\temp
\def\hour{\the\hours}
\def\minute{\ifnum\minutes<10 0\the\minutes
            \else\the\minutes\fi}
\def\clock{
\ifnum\hours=0 12:\minute\ AM
\else\ifnum\hours<12 \hour:\minute\ AM
      \else\ifnum\hours=12 12:\minute\ PM
            \else\ifnum\hours>12
                 \pmhours=\hours
                 \advance\pmhours by -12
                 \the\pmhours:\minute\ PM
                 \fi
            \fi
      \fi
\fi
}

\def\monthname{\relax\ifcase\month 0/\or January\or February\or
   March\or April\or May\or June\or July\or August\or September\or
   October\or November\or December\else\number\month/\fi}

\def\bold#1{\setbox0=\hbox{$#1$}%
     \kern-.025em\copy0\kern-\wd0
     \kern.05em\copy0\kern-\wd0
     \kern-.025em\raise.0433em\box0 }

\def\beq{\begin{equation}}
\def\eeq{\end{equation}}

\def\ga{\mathrel{\raise.3ex\hbox{$>$\kern-.75em\lower1ex\hbox{$\sim$}}}}
\def\la{\mathrel{\raise.3ex\hbox{$<$\kern-.75em\lower1ex\hbox{$\sim$}}}}
\def\gev{{\rm \, Ge\kern-0.125em V}}
\def\tev{{\rm \, Te\kern-0.125em V}}
\def\gyr{{\rm \, G\kern-0.125em yr}}

\def\gappeq{\mathrel{\rlap {\raise.5ex\hbox{$>$}}
{\lower.5ex\hbox{$\sim$}}}}
\def\lappeq{\mathrel{\rlap{\raise.5ex\hbox{$<$}}
{\lower.5ex\hbox{$\sim$}}}}
\def\Toprel#1\over#2{\mathrel{\mathop{#2}\limits^{#1}}}

\def\sl{{\widetilde \ell}_{\scriptscriptstyle\rm R}}

\def\m12{m_{1\!/2}}

\def\eg{{e.g. }}
\def\ie{{i.e. }}

\def\et{{{\it et al.}}}
\def\etc{{etc.}}

\def\'{^{\prime}}

\def\avrg#1{{\langle #1 \rangle}}

\def\hmpc{{\, {\rm h}^{-1}~\rm Mpc}}

\def\kpc{{\rm~kpc}}
\def\mpc{{\rm~Mpc}}

\def\spose#1{\hbox to 0pt{#1\hss}}
\def\lta{\mathrel{\spose{\lower 3pt\hbox{$\mathchar"218$}}
     \raise 2.0pt\hbox{$\mathchar"13C$}}}
\def\gta{\mathrel{\spose{\lower 3pt\hbox{$\mathchar"218$}}
     \raise 2.0pt\hbox{$\mathchar"13E$}}}
\def\ge{\mathrel{\spose{\lower 3pt\hbox{$-$}}
     \raise 2.0pt\hbox{$\mathchar"13E$}}}
\def\le{\mathrel{\spose{\lower 3pt\hbox{$-$}}
     \raise 2.0pt\hbox{$\mathchar"13C$}}}

\newcommand{\DASI}{DASI}

\def\ca{{{\it ca.}}}
\def\cf{{compared with}}
\def\janzerozero{{January 2000}}
\def\janzerotwo{{January 2002}}
\def\junzerotwo{{June 2002}}
\def\janzerothree{{January 2003}}
\def\marchzerothree{{March 2003}}

\def\Acbar{{ACBAR}}
\def\Archeops{{ARCHEOPS}}

\def\npwmapS{{Spergel \et\ 2003}}
\def\npwmapV{{Verde \et\ 2003}}
\def\npKuo02{{Kuo \et\ 2003}}
\def\npGoldstein02{{Goldstein \et\ 2003}}
\def\npcapp00{{Bond \et\ 2000}}
\def\npbmrst02{{Bond \et\ 2003a}}

\def\npbjkrad00{{Bond, Jaffe \& Knox 2000}}
\def\npbjkquad98{{Bond, Jaffe \& Knox 1998}}
\def\npbj98{{Bond \& Jaffe 1999}}
\def\nptoco99{{Miller \et\ 1999}}
\def\npmauskopf99{{Mauskopf \et\ 2000}}
\def\npdeBnature00{{de Bernardis \et\ 2000}}
\def\nplange00{{Lange \et\ 2001}}  
\def\npjaffe00{{Jaffe \et\ 2001}}  
\def\npmaxhanany00{{Hanany \et\ 2000}}
\def\npnett01{{Netterfield \et\ 2002}}
\def\npdeBpkdip01{{de Bernardis \et\ 2002}}
\def\npmaxlee01{{Lee \et\ 2002}}
\def\npRuhl02{{Ruhl \et\ 2003}}
\def\npDASI01{{Halverson \et\ 2002}}
\def\npMason02b{{Mason \et\ 2003}}
\def\npPearson02{{Pearson \et\ 2003}}
\def\npMyers02{{Myers \et\ 2003}}
\def\npSievers02{{Sievers \et\ 2003}}
\def\npBond02{{Bond \et\ 2003b}}
\def\npReadhead03{{Readhead \et\ 2003}}
\def\npcmbfast{{Seljak \& Zaldarriaga 1996}}
\def\npcamb{{Lewis, Challinor \& Lasenby 2000}}
\def\npcosmomc{{Lewis \& Bridle 2002}}
\def\npmetropolis{{Metropolis \et\ 1953}}
\def\npMCMCa{{Christensen \& Meyer 2001}}
\def\npMCMCb{{Christensen \et\ 2001}}
\def\npkowsowsky02{{Kosowsky \et\ 2002}}
\def\npDASIpol02{{Leitch \et 2002, Kovac \et\ 2002}}
\def\npBIMA02{{Dawson \et\ 2002}}
\def\npbc01{{Bond \& Crittenden 2001}}
\def\npbh95{{Bond 1996}}
\def\npdegeneracies{{Efstathiou \&  Bond 1999}}
\def\npSN{{Riess \et 1998, Perlmutter \et\ 1999a,b, 2003}}
\def\npFreedman01{{Freedman \et\ 2001}}
\def\npbbn03{{Kirkman \et\ 2003}}

\def\npCKK03{{Chu, Kaplinghat \& Knox 2003}}
\def\npVSA02{{Scott \et\ 2002}}
\def\npVSAext02{{Grainge \et\ 2003}}
\def\npBPS00top{{Bond, Pogosyan \&  Souradeep 2000}}
\def\nparcheops02{{Benoit \et\ 2003}}

\def\wmapB{{Bennett \et\ (2003)}}
\def\wmapH{{Hinshaw \et\ (2003)}}

\def\wmapP{{Page \et\ (2003)}}
\def\wmapS{{Spergel \et\ (2003)}}
\def\wmapV{{Verde \et\ (2003)}}
\def\Kuo02{{Kuo \et\ (2003)}}
\def\Goldstein02{{Goldstein \et\ (2003)}}
\def\capp00{{Bond \et\ (2000)}}
\def\bmrst02{{Bond \et\ (2003a)}}

\def\bjkrad00{{Bond, Jaffe \& Knox (2000)}}
\def\bjkquad98{{Bond, Jaffe \& Knox (1998)}}
\def\bj98{{Bond \& Jaffe (1999)}}
\def\toco99{{Miller \et\ (1999)}}
\def\mauskopf99{{Mauskopf \et\ (2000)}}
\def\deBnature00{{de Bernardis \et\ (2000)}}
\def\lange00{{Lange \et\ (2001)}}  
\def\jaffe00{{Jaffe \et\ (2001)}}  
\def\maxhanany00{{Hanany \et\ (2000)}}
\def\nett01{{Netterfield \et\ (2002)}}
\def\deBpkdip01{{de Bernardis \et\ (2002)}}
\def\maxlee01{{Lee \et\ (2002)}}
\def\Ruhl02{{Ruhl \et\ (2003)}}
\def\DASI01{{Halverson \et\ (2002)}}
\def\Mason02b{{Mason \et\ (2003)}}
\def\Pearson02{{Pearson \et\ (2003)}}
\def\Myers02{{Myers \et\ (2003)}}
\def\Sievers02{{Sievers \et\ (2003)}}
\def\Bond02{{Bond \et\ (2003b)}}
\def\Readhead03{{Readhead \et\ (2003)}}

\def\kowsowsky02{{Kosowsky \et\ (2002)}}
\def\DASIpol02{{Leitch \et (2002), Kovac \et\ (2002)}}
\def\BIMA02{{Dawson \et\ (2002)}}
\def\bc01{{Bond \& Crittenden (2001)}}
\def\bh95{{Bond (1996)}}

\def\Freedman01{{Freedman \et\ (2001)}}
\def\bbn03{{Kirkman \et\ (2003)}}

\def\CKK03{{Chu, Kaplinghat \& Knox (2003)}}
\def\VSA02{{Scott \et\ (2002)}}
\def\VSAext02{{Grainge \et\ (2003)}}
\def\BPS00top{{Bond, Pogosyan \&  Souradeep (2000)}}
\def\archeops02{{Benoit \et\ (2003)}}

\begin{document}

\title[CMB snapshots: pre-WMAP and post-WMAP]{Cosmic microwave background snapshots: pre-WMAP and post-WMAP}
\author[J.R. Bond, C.R. Contaldi, D. Pogosyan]{J. Richard Bond$^{1,2,4}$, Carlo Contaldi$^{1}$, Dmitry
Pogosyan$^{3}$}
\affiliation{1. Canadian Institute for Advanced Research Cosmology \& Gravity 
Program, \\ Canadian Institute for Theoretical Astrophysics, McLennan Physical Laboratories, University of Toronto, 60 St. George Street, Toronto, 
Ontario M5S 3H8, Canada \\ 2. Institut d'Astrophysique de Paris, 98bis Boulevard Arago, 75014 Paris, France \\
3. Department of Physics, University of Alberta,\\
412 Avadh Bhatia Physics Laboratory, Edmonton, Alberta T6G 2J1, Canada \\ 4. Institute of Astronomy, Madingley Road, Cambridge CB3 0HA, UK \\[2ex]}

\maketitle
\begin{abstract}{cosmology, particle physics, dark matter, dark energy}
\noindent
{\small {\it Phil.~Trans.~R.~Soc.~Lond.~A} {\rm {\bf 361}, 2435-2467 (2003).} {\it Published online October 15, 2003.}}
\vskip 4pt
We highlight the remarkable evolution in the cosmic microwave
background (CMB) power spectrum ${\cal C}_\ell$ as a function of
multipole $\ell$ over the past few years, and in the cosmological
parameters for minimal inflation models derived from it: from
anisotropy results before 2000; in 2000, 2001, from Boomerang, Maxima
and the Degree Angular Scale Interferometer (DASI), extending $\ell$
to approximately $1000$; and in 2002, from the Cosmic Background
Imager (CBI), the Very Small Array (VSA), \Archeops\ and the Arcminute
Cosmology Bolometer Array Receiver (\Acbar), extending $\ell$ to
approximately $3000$, with more from Boomerang and DASI as
well. Pre-WMAP (pre-Wilkinson Microwave Anisotropy Probe) optimal
bandpowers are in good agreement with each other and with the
exquisite one-year WMAP results unveiled in February 2003, which now
dominate the $\ell \lta 600$ bands. These CMB experiments
significantly increased the case for accelerated expansion in the
early universe (the inflationary paradigm) and at the current epoch
(dark energy dominance) when they were combined with `prior'
probabilities on the parameters.

The minimal inflation parameter set,
$\{\omega_b,\omega_{cdm},\Omega_{tot}, \Omega_\Lambda,n_s,\tau_C,
\sigma_8\}$, is applied in the same way to the evolving data. ${\cal
C}_\ell$ database and Monte Carlo Markov Chain methods are shown to
give similar values, highly stable over time and for different prior
choices, with the increasing precision best characterized by
decreasing errors on uncorrelated `parameter eigenmodes'.  Priors
applied range from weak ones to stronger constraints from the
expansion rate (HST-h), from cosmic acceleration from supernovae (SN1)
and from galaxy clustering, gravitational lensing and local cluster
abundance (LSS).  After marginalizing over the other cosmic and
experimental variables for the weak+LSS prior, the pre-WMAP data of
\janzerothree\  \cf\  the post-WMAP data of \marchzerothree\  give $\Omega_{tot}
=1.03^{+0.05}_{-0.04}$ \cf\  $1.02^{+0.04}_{-0.03}$, consistent with
(non-baroque) inflation theory. Adding the flat $\Omega_{tot}=1$
prior, we find a nearly scale invariant spectrum, $n_s
=0.95^{+0.07}_{-0.04}$ \cf\ $0.97^{+0.02}_{-0.02}$.  The evidence
for a logarithmic variation of the spectral tilt is $\lta 2\sigma$. 
The densities are: for baryons,
$\omega_b\equiv \Omega_b {\rm h}^2 =0.0217^{+0.002}_{-0.002}$ \cf\ $
0.0228^{+0.001}_{-0.001}$, near the the Big Bang nucleosynthesis estimate
of $0.0214\pm 0.002$; for CDM, $\omega_{cdm}=\Omega_{cdm}{\rm h}^2
=0.126^{+0.012}_{-0.012}$ \cf\ $ 0.121^{+0.010}_{-0.010}$; and for the
substantial dark (unclustered) energy, $\Omega_\Lambda \approx
0.66^{+0.07}_{-0.09}$ \cf\  $0.70^{+0.05}_{-0.05}$. The dark energy
pressure-to-density ratio $w_Q$ is not well constrained by our
weak+LSS prior, but adding SN1 gives $w_Q \lta -0.7$ for \janzerothree\  and
\marchzerothree, consistent with the $w_Q$=$-1$ cosmological constant case. We
find $\sigma_8 = 0.89^{+0.06}_{-0.07}$ \cf\ $0.86^{+0.04}_{-0.04}$,
implying a sizable Sunyaev-Zeldovich (SZ) effect from clusters and groups; the high $\ell$
power found in the \janzerothree\  data suggest $\sigma_8 \sim
0.94^{+0.08}_{-0.16}$ is needed to be SZ-compatible.
\end{abstract}

\section{The evolution of CMB spectra and cosmic parameters} \label{sec:GUSparam}

We have been in the midst of a remarkable outpouring of results from
the CMB since 1999. The Royal Society Discussion Meeting focused on
the eight pre-Wilkinson Microwave Anisotropy Probe (pre-WMAP)
announcements made in 2002. The WMAP release was three weeks later,
and a before-WMAP discussion without an after-WMAP discussion is
unthinkable now. This paper applies the same methods of analysis to
WMAP as to the earlier CMB experiments to put its singular forward
step into context. In \S~\ref{sec:cmbdata}, we describe the different
experiments that have contributed to the evolving picture. For more
background material on methods and references, see \bh95, \bc01,
Bond \et\ (2003a,b), \Sievers02, \Ruhl02\ and \Goldstein02.

\subsection{Grand unified spectra compared with WMAP} \label{sec:GUS}

\begin{figure}[h]
\begin{center}
\includegraphics[width=5.0in]{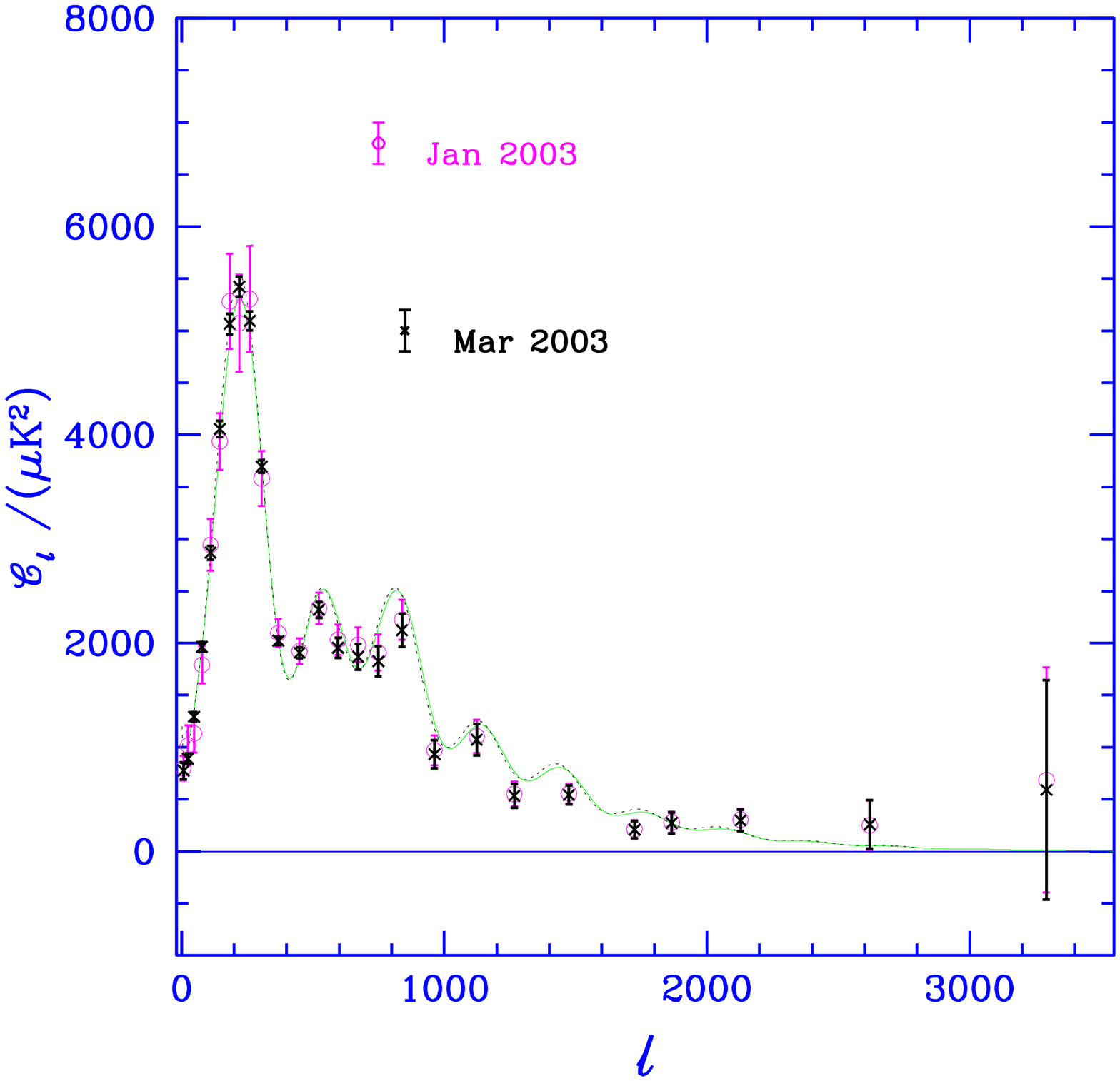}
\end{center}
\caption{Optimal ${\cal C}_\ell$ spectra for the pre-WMAP \janzerothree\  data
and the post-WMAP \marchzerothree\  data show good agreement.  These spectra are
maximum-likelihood determinations of the power in 26 (top-hat) bands,
with calibration and beam uncertainties of the various experiments
fully taken into account. Two ${\cal C}_\ell$ $\Lambda$CDM models from
the database are shown. The dotted (black) curve best fits the \marchzerothree\ 
data with the weak+flat+LSS+$\tau_C$ prior applied. It has parameters
$\{\Omega_{\rm {tot}}$, $\Omega_\Lambda$, $\Omega_b h^2$, $\Omega_{\rm
cdm} h^2$, $n_s$, $\tau_C, t_0, h, \sigma_8\}$ = $\{1.0, 0.7, 0.0225,
0.12, 0.975, 0.15, 13.7, 0.69, 0.89 \}$. The solid green curve that
looks quite similar best fits the \junzerotwo\  data for the weak+flat+LSS
prior (Sievers \et\ 2003), with parameters $\{1.0, 0.5, 0.020, 0.14,
0.925, 0, 14.4, 0.57, 0.82\}$. It was used as the inter-band shape for
this optimal bandpower determination, but the results are insensitive
to this.  The bandpowers are optimally placed in $\ell$. Their finite
horizontal extension is not shown, and the vertical diagonal bandpower
errors also do not show the whole story since there are band-to-band
correlations. (\eg the visual up-down-up at the first peak for \janzerothree\ 
is indicative of the strong correlations and can disappear with different
banding, \eg one better tuned to Boomerang's binning.) Despite these
caveats, the best fit ${\cal C}_\ell$ would fit better with a slight
downward tilt beyond $\ell \gta 500$, which a scale-dependent $n_s(k)$
could do (see \S~\ref{sec:params}).  }
\label{fig:CLoptmar03}
\end{figure}

\begin{figure}
\begin{center}
\includegraphics[width=5.0in]{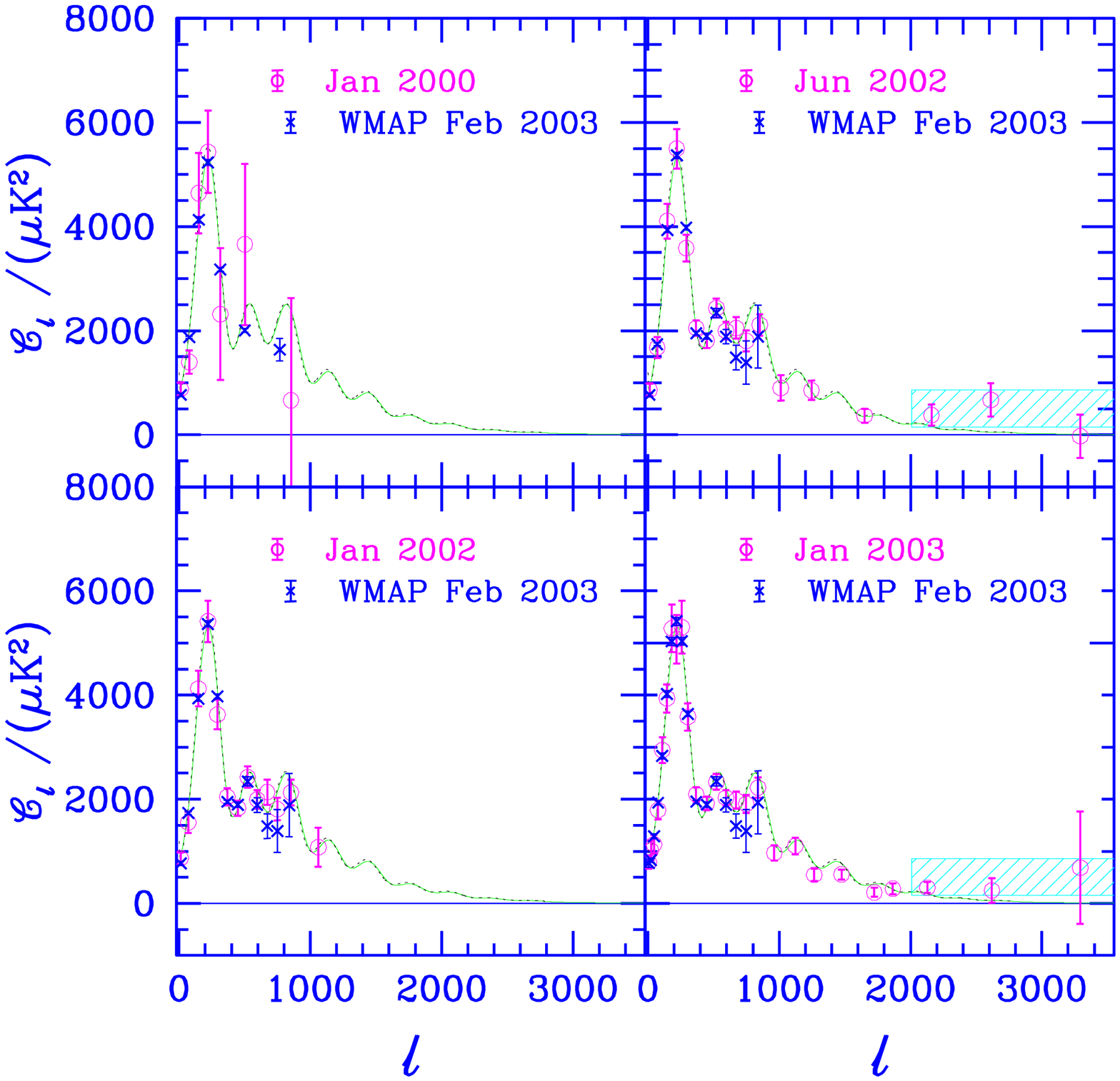}
\end{center}
\caption{The evolving optimal ${\cal C}_\ell$ values are compared with
WMAP-only power spectra (crosses) compressed onto the same bands.  The
degree of compression does not do visual justice to WMAP because the
errors are so small until beyond the second peak. The bands were
chosen to be natural for the data at the time, but band-to-band
correlations do exist. The spectra include the following data: \janzerozero\ 
has DMR + Toco + Boomerang-NA + the April 1999 mix; \janzerotwo\  has \janzerozero\ 
and the Boomerang data of Netterfield \et\ (2001) + Maxima + DASI;
\junzerotwo\  has \janzerotwo\  plus one-year CBI mosaic and deep field data and the
(non-extended) VSA data; \janzerothree\  has \Archeops\ + \Acbar, uses the Ruhl \et\ 
(2003) Boomerang spectrum covering 2.9\% of the sky, the extended-VSA
data and the two-year combined CBI mosaic + deep field data. The $\ell >
2000$ excess found with the one-year deep CBI data is denoted by the light blue
hatched region (95\% confidence limit) in the right hand panels. The
two best fit $\Lambda$CDM models of figure~\ref{fig:CLoptmar03} are
repeated in each of the panels. When HST-h or SN1 priors are included
in the \junzerotwo\  data, the best fit model has the same parameters as 
those of the \marchzerothree\  curve, except for a slight shift in tilt, to
$n_s$=1.0, a corresponding rise in $\tau_C$, to 0.20, leading to
$\sigma_8$=0.91. }
\label{fig:CLoptevoln}
\end{figure}

Figure~\ref{fig:CLoptmar03} shows how the CMB power spectrum, ${\cal
C}_\ell \equiv \ell (\ell+1)\avrg{\vert T_{\ell m}\vert^2}/(2\pi) $
defined in terms of CMB temperature anisotropy multipoles $T_{\ell
m}$, changed from the pre-WMAP determination shown at the Royal
Society meeting using the data to January 2003 to the post-WMAP
determination of March 2003; the agreement is excellent. How we got
there is shown in figure~\ref{fig:CLoptevoln}, providing snapshots
compared with WMAP for data that was available in \janzerozero, \janzerotwo, \junzerotwo,
as well as \janzerothree.  Accompanying this story is a convergence with
decreasing errors over time on the values of the cosmological
parameters given in figures~\ref{fig:2Djan000203mar03},
\ref{fig:1DwkflatLSS} and table~\ref{tab:exptparams}.

Optimal spectra and their error matrices are calculated in exactly the
same way that cosmological parameters are, with the parameters now the
bandpowers ${\cal C}_{\rm b}$ in the chosen $\ell$-bins, ${\rm b}$. Additionally,
characterizing each experiment there are calibration uncertainties and
often beam uncertainties, each adding additional parameters. Sample
values for these are given in \S~\ref{sec:analysis}\ref{sec:calib}.

\subsection{Basic parameters characterizing the early Universe and CMB transport} \label{sec:CMBparams}

\begin{figure}
\includegraphics[width=5.2in]{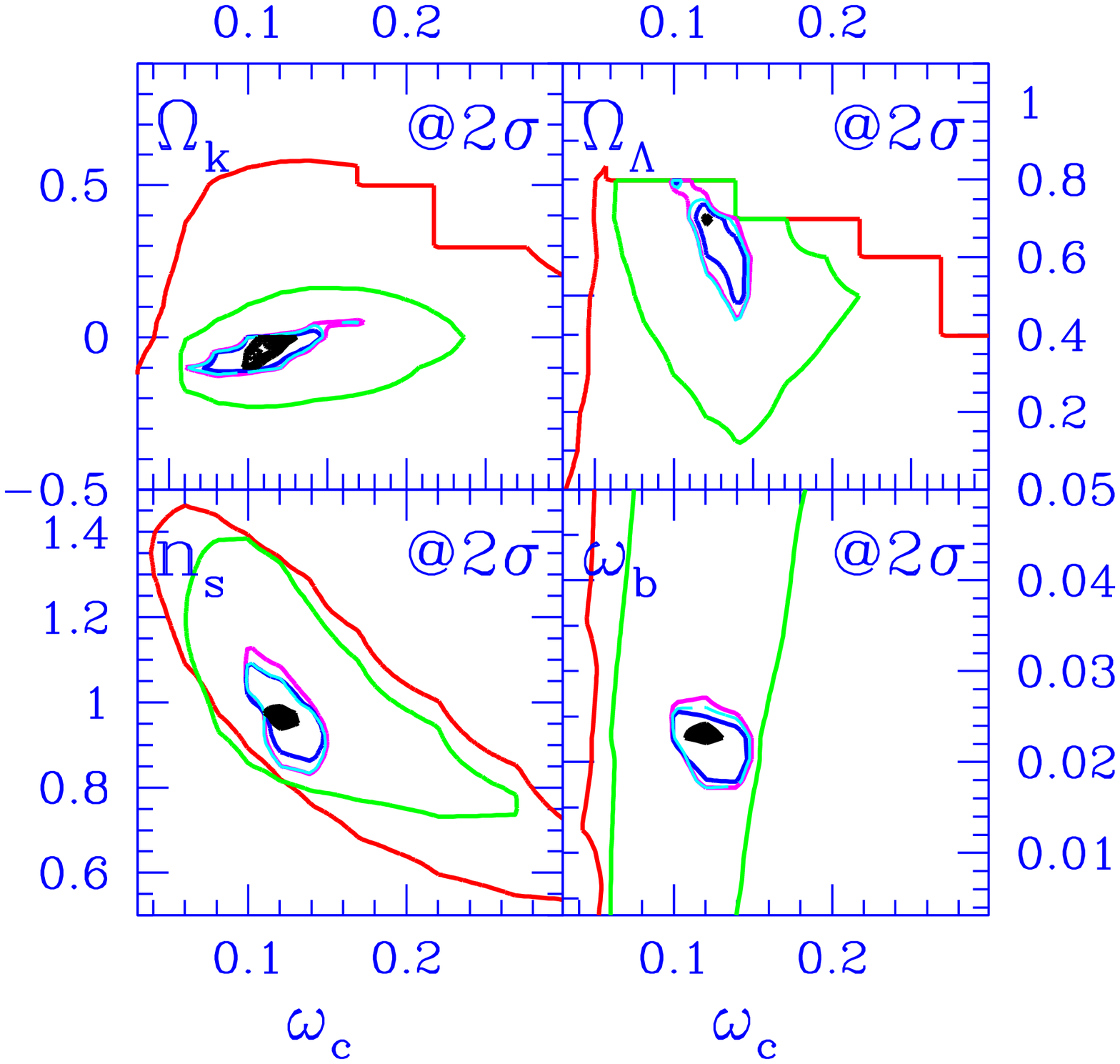}
\caption{The evolution of 2$\sigma$ likelihood-contour regions, with the
weak+LSS prior probability applied in the top-left-hand panel, and the flat
$\Omega_{tot}$=1 applied additionally in the rest. The outer (red) contour
is for COBE-DMR only, then the sequence of increasing concentration is
for the \janzerozero\  (green), \janzerotwo\  (magenta), \junzerotwo\  (cyan) and \janzerothree\  (blue)
data. The dense (black) region is for \marchzerothree, including WMAP, but
excluding \Archeops\ and DMR because of their substantial overlap with
WMAP. A $\tau_C$ prior (see \S~\ref{sec:GUSparam}\ref{sec:prior}) taking
into account the WMAP `model independent' TE analysis, has been
included, but it does not make much difference to these results. If
only the weak prior is imposed, the \janzerozero\  data still show rough
$\Omega_k$ localization, but do not constrain the other
parameters significantly, whereas the \janzerotwo\  and \janzerothree\  contour regions
without LSS are  only slightly bigger than those shown here. For \marchzerothree, the black
regions remain small, but the $\Omega_\Lambda -\Omega_k$ degeneracy
becomes more evident, with closed universes and smaller
$\Omega_\Lambda$ allowed. }
\label{fig:2Djan000203mar03}
\end{figure}

Our philosophy has been to consider minimal models first, then see how
progressive relaxation of the constraints on the inflation models, at
the expense of increasing baroqueness, causes the parameter errors to
open up. We adopt the basic set of seven cosmological parameters
$\{\omega_b,\omega_{cdm}, \Omega_\Lambda,\Omega_{k},
n_s,\tau_C,\ln{\cal P}_{\Phi}(k_n) \}$ to facilitate comparison with
results in \lange00, \jaffe00, \nett01, \Sievers02 and \Goldstein02. How
the values have converged upon the bull's-eye $2\sigma$ determinations
with WMAP is shown in figure~\ref{fig:2Djan000203mar03}. In spite of the
great success in extending the spectrum to high $\ell$, the evolution
of the parameter errors was not that strong after \janzerotwo\  until
WMAP. This is because the ${\cal C}_\ell$ model space is restrictive
for inflation-based models, with high $\ell$ intimately related to
lower $\ell$. On the other hand, when the experiments were treated
individually (always with COBE-DMR), their $2\sigma$ contours were all
circling the bull's-eyes (Sievers \et\ 2003).

The transport of the radiation through the era of photon decoupling 
is sensitive to the physical density of all of the species of
particles present then, $\omega_j \equiv \Omega_j {\rm h}^2$. We use
two parameters, $\omega_b$ for baryons and $\omega_{cdm}$ for cold
dark matter, to characterize this, but we should add $\omega_{hdm}$
for hot dark matter (massive but light neutrinos), and $\omega_{er}$
for the relativistic particles present at that time (photons, very
light neutrinos, and possibly weakly interacting products of late time
particle decays). Here the latter is fixed for the conventional three 
species of relativistic neutrinos plus photons.  The total matter
density is 
\begin{equation} 
\omega_m = \omega_b + \omega_{cdm} +\omega_{hdm} \, . \nonumber 
\end{equation} 
Another two parameters characterize the transport from decoupling to
the present, the vacuum or dark energy, encoded in a cosmological
constant $\Omega_\Lambda$, and the curvature energy $\Omega_k \equiv
1-\Omega_{tot}$. Of course $\Omega_k$ also determines the mean
geometry. (When one wishes to focus on what the CMB can tell us about
the nature of the dark energy, another parameter is often added, $w_Q
= p_Q /\rho_Q$, where $p_Q$ and $\rho_Q$ are the pressure and density
of the dark energy. If the vacuum or dark energy is reinterpreted as
$\Omega_Q$, the energy in a scalar field $Q$ which dominates at late
times, it would be likely to have complex dynamics associated with
it. In that case, $Q$ and $w_Q$ would have spatial and temporal
variations (except if $w_Q=-1$, the cosmological constant
case). Spatial fluctuations of $Q$ are expected to leave a direct
imprint on the CMB for small $\ell$. This complication is typically
ignored, but should not be. It does depend in detail upon the specific
model for $Q$.)

In this parameter space, ${\rm h}= (\sum_j \omega_j
)^{1/2}$ and the age of the Universe $t_0$ are derived functions
of the $\omega_j $, $\Omega_{k,\Lambda}$ and $w_Q$.

Another parameter is the Compton `optical depth' $\tau_C$ from a
reionization redshift $z_{reh}$ to the present, 
\begin{equation} 
\tau_C \approx 0.12
(\omega_b/0.02) (\omega_m/0.15)^{-1/2} ((1+z_{{\rm reh}})/15
)^{3/2} \, . \nonumber 
\end{equation} 
As long as $\tau_C$ is not too large, ${\cal C}_\ell$ is
suppressed by a factor $\exp[-2\tau_C]$ on scales smaller than the
horizon at $z_{\rm reh}$. For typical models of hierarchical structure
formation, we expect $\tau_C \lta 0.3$. At the moment, even with WMAP,
the CMB total anisotropy (TT) alone does not give such a
constraint. It is the cross correlation of total anisotropy with
polarization (TE) that leads to a detection (Kogut \et\ 2003).

Two parameters characterize the early universe primordial power
spectrum of gravitational potential fluctuations $\Phi$, one giving
the overall power spectrum amplitude ${\cal P}_{\Phi}(k_n)$, and one
defining the shape, a spectral tilt 
\begin{equation} 
n_s (k_n) \equiv 1+d\ln {\cal
P}_{\Phi}/d \ln k \, ,  \nonumber 
\end{equation} 
at some (comoving) normalization wavenumber
$k_n$. Instead of $\ln {\cal P}_\Phi (k_n)$, which is appropriate for
connecting to early universe physics, we use as a basic amplitude
variable $\ln {\cal C}_{10}$ when connecting to CMB, and $\ln
\sigma_8^2$ when connecting to LSS.

To characterize inflation, even in the simplest models, we really need
at least another two parameters, ${\cal P}_{GW}(k_n)$ and $n_t(k_n)$,
associated with the gravitational wave (GW) component. In inflation,
the amplitude ratio ${\cal P}_{GW}/{\cal P}_{\Phi}$ is related to
$n_t$ to lowest order, with ${\cal O}(n_s-n_t)$ corrections at higher
order (\eg \npbh95). There are also useful limiting cases for the
$n_s-n_t$ relation.  With \janzerothree\  data, and even with WMAP, the data
are not powerful enough to determine much about the GW contribution,  
\eg the WMAP team estimate the gravitational wave (tensor)
contribution to be less than $0.72$ of the scalar component in amplitude at
the 95\% CL.

As one allows the baroqueness of the inflation models to increase, one
can entertain essentially any power spectrum. This implies a fully
$k$-dependent $n_s(k)$ if one is artful enough in designing inflaton
potential surfaces. The simple model 
\begin{equation} 
n_s(k) = n_s (k_n) +
[dn_s(k_n)/d\ln k] \, \ln (k/k_n)\,   \nonumber 
\end{equation} 
adds a logarithmic running index about
a pivot scale $k_n$. As figure~\ref{fig:CLoptmar03} and
\S~\ref{sec:params} indicate, this improves the fit to the data.
It is also expected in inflation models, it is just a question of the
size of the correction.

The tensor index $n_t(k)$ could also be a function, although it does
not have as much freedom as $n_s(k)$ in inflation. For example, it is
difficult to get $n_t(k)$ to be positive. One can also have more
types of modes present, \eg scalar isocurvature modes. The data 
have shown for quite a while  that these would have to be subdominant relative to the
scalar curvature modes, and would have to be even more so now.

Each experiment also contributes a parameter describing the
uncertainty in the calibration, and possibly another for the
uncertainty in the beam size.

\subsection{CMB analysis pipelines: bandpowers to cosmic parameters} \label{sec:bandtoparams}

In Gaussian models defined by a parameter set $\{ y_a \}$, the
probability distribution of the primary anisotropies is fully encoded
in the isotropic power spectrum ${\cal C}_\ell (y_a)$ -- as long as
there is no preferred orientation (as might occur for small universes
that are topologically nontrivial). The observed bandpowers for an
individual experiment can then be tested against theoretical
bandpowers ${\cal C}_{\rm b} (y_a)$, which are averages of the ${\cal
C}_\ell$'s over $\ell$-space `window functions' $\varphi_{{\rm b}\ell}$
appropriate to the bands for the experiment in question. This
represents a huge compression of the entire dataset and makes large
model space computations feasible.

 To use this information to estimate cosmological parameters, the
entire likelihood surface as a function of the $\{ {\cal C}_{\rm b} \}$ is
needed with sufficient accuracy that the parameter estimations are not
biased. It has been shown that individual bandpowers ${\cal C}_{\rm b}$ have
distributions well characterized by a lognormal distribution in the
variable ${\cal C}_{\rm b} + {\cal C}_{N{\rm b}}$, where ${\cal C}_{N{\rm b}}$ is an
estimate of the noise in the band (\npbjkrad00).  The coupling between
bandpowers is included as a weak correction, relying on the
band-to-band correlations being relatively small --- a demand imposed
in the data analysis phase. What comes out are entropies, $S(y_a)$,
\ie log likelihoods. A slightly modified version of this
prescription is used for WMAP (\wmapV; see also \S~\ref{sec:cmbdata}\ref{sec:mar03}.)

There are two approaches to sampling the set $\{ y_a \}$ that we have
used.  The main workhorse throughout our analyses up to \janzerothree\  used
fixed grids: a discrete set of parameter values are chosen {\it a
priori} for six of the seven cosmic variables, with spacings in each of the
dimensions designed by hand to be adaptively concentrated about the
most probable values, but with sufficient spread to ensure that tails
and multiple solution regions are well explored.  The current database
for the `minimal inflation' parameter model contains $8.5 \times
10^6$ models, with dimensions $15 \times 13 \times 15 \times 12 \times
31 \times 11$ for the `external parameter' set
$\{\omega_b,\omega_{cdm},\Omega_{tot}, \Omega_\Lambda,n_s,\tau_C\}$,
with edge cutouts requiring $\Omega_m \ge 0.1$. 

The seventh (amplitude) parameter, $\ln{\cal C}_{10}$ or $\ln
\sigma_8^2$, and the experimental calibration and beam uncertainty
variables are continuous. They relax to their maximum-likelihood
values, with errors characterized by the second derivative of the
likelihood function. The number of these `internal' continuous
parameters may become much larger if we split the amplitude parameter
into many, one for each band in $\ell$-space (or $k$-space if
three-dimensional power spectra are the target). The shape ${\cal
C}^{(s)}_\ell$ of an assumed spectrum multiplies the adopted window
functions for the bands. For the optimal bandpowers that combine
experiments together in figure~\ref{fig:CLoptevoln}, ${\cal
C}^{(s)}_\ell$ is usually varied to test robustness of the results,
but an ensemble of external parameter models can be applied, \eg\ in
broken scale-invariance applications in $k$-space.

The first stage output is large entropy files that include
maximum-likelihood values and Fisher matrices for the internal
variables. These files are then picked up in postprocessing as various
prior probabilities are applied, marginalizations are done, and
one-dimensional (1D) and two-dimensional (2D) statistics are computed.

An advantage of a fixed grid is that it has allowed us to quickly
check many prior cases and many experimental combinations, all on the
same footing. Calibration uncertainties are handled either at the
entropy stage with the complete experimental mix, or in
postprocessing, since we know the amplitude distributions. When
analysing an experiment, these operations are done again and again,  as
different hypotheses, band widths and positionings, estimation
techniques, source removal methods, \etc, are applied to the
bandpowers for the experiments in question, so speed is 
essential. Combining DMR with the experiments was always a first
step; now WMAP takes that role.

As new experiments are added which are qualitative improvements (like
WMAP), errors may become smaller than grid spacings and further
adaptivity of the grid is needed. This is so even with good
interpolations and smoothing. We set a floor on parameter errors to be
half the grid spacing of the encompassed grid: a value that was never
reached before WMAP but has been with WMAP for a few parameters strongly
bundled into the top few parameter eigenmodes and some priors.

The second approach is the Monte Carlo Markov Chain (MCMC) method
(\npmetropolis; \npMCMCa; \npMCMCb; \npcosmomc; \npwmapV). It
develops a set of independent chains, each a small (unstructured) grid
on the parameter space that is constructed `on the fly' rather than
{\it a priori}.  The elements of the chains are sampled according to
well-developed MCMC algorithms designed to make the next step
independent of prior ones. The spacing of the models computed changes
with experimental combinations and priors adopted. As with the
fixed-grid methods, some priors can be applied in postprocessing,
which speeds up the procedure. These MCMC methods have now become
feasible for CMB analyses because the ${\cal C}_\ell (y_a) $
computations with CMBfast (\npcmbfast) or CAMB (\npcamb) are efficient
algorithmically, and are rapid enough on large numbers of models
because of the remarkable speedup of individual computer processors in
recent years. A nice Fortran 90 package is publicly available to do
this (\npcosmomc). To do the statistics well, one needs not just many
elements in a chain, but a number of chains.  Thus it was really the
advent of massively parallel machines that is allowing MCMC to become
a major working tool for repetitive CMB and LSS analyses. For example,
it is the method adopted by the WMAP team (\npwmapV; \npwmapS), who
applied it to the minimal six-parameter flat model, with the amplitude
as an external parameter, and seven-parameter models with the
$\Omega_k$, $w_Q$, $dn_s/d\ln k$ or $n_t$ allowed to vary in
turn. They ran four chains and about 30000 models per chain to define
their distributions. We have adapted Cosmomc to the parameter choices
and ranges of our ${\cal C}_\ell$ database, and the many prior cases
and experimental combinations used, to facilitate comparisons; \eg
$\tau_C$ can go out as far as 0.7, which ensures likelihood drop-off
to zero, but places sampling challenges. The major challenge for MCMC
is to sample well the curved likelihood ridges at reasonable
computational cost.  Certain nonlinear combinations of our basic
variables can help to straighten out the likelihood surfaces, in
particular those with highly asymmetric errors, allow for more
efficient and accurate computation, whether they be used in MCMC
(\npkowsowsky02; \npCKK03; \npwmapV) or for fixed or adaptive
grids. For MCMC, another approach is to use variance matrices from
small runs to make the steps efficient in the parameter space
(\npcosmomc). For each experimental mix and prior, we use 16 chains
run until convergence tests are satisfied for all of the variables.
In spite of processor speed, the computations remain a challenge if
many cases need to be run.

\subsection{Weak, HST-h, SN1, LSS \& $\tau_C$ priors} \label{sec:prior}

The parameter grids are chosen to be wide relative to conceivable
cosmological models, yet are concentrated in the maximum-likelihood
regions. The MCMC chains are allowed to vary over wide domains, and
they automatically concentrate well. An important issue is the prior
measure we impose upon the parameter space. Implicit in the adoption
of a given variable set is that a uniform prior probability is chosen
in each of the variables. If a variable is not well determined this
can have a big influence (\nplange00).

We usually present the cosmological conclusions we draw from our
analyses of the various CMB experiments using noncontroversial priors,
ones that almost all cosmologists would agree to. Thus our standard
weak one used in Lange \et\ (2001) and subsequent works requires only
$0.45 \le {\rm h} \le 0.90$ and $t_0 \ge 10 \, {\rm Gyr}$. The addition of the
flat prior has also become benign, thanks to the sharpness of the
$\Omega_k \approx 0 $ determination with the CMB rather than to the
predilections of inflation theorists. (Although a major reduction in
number of database models occurs when the flat $\Omega_k=0$ prior is
applied, it is usually applied in the postprocessing phase.)

Data from sources other than the CMB can be incorporated as `prior
probabilities'. A stronger prior on the Hubble parameter, HST-h, uses
an $h=0.72 \pm 0.08$ Gaussian distribution (\npFreedman01). SN1 data
imposes a prior in $\Omega_\Lambda$-$\Omega_k$-$w_Q$ space (\npSN). The
CMB data apparently determine $\omega_b$ to higher accuracy than
light-element-abundance observations coupled to Big Bang
Nucleosynthesis theory (\npbbn03), hence applying a BBN prior 
is not of much interest.

The LSS prior we use (\npbj98; \nplange00; \npBond02) also depends upon our
parameter set. An important combination is the wavenumber of the
horizon when the energy density in relativistic particles equals the
energy density in nonrelativistic particles: $k_{Heq}^{-1} \approx 5
\Gamma_{eq}^{-1} \hmpc$, where $\Gamma_{eq} = \Omega_m {\rm h} \ (1.68
\omega_\gamma /\omega_{er})^{1/2}$.  We represent the (linear) density power
spectrum by a single shape parameter:
\begin{equation} 
\Gamma = \Gamma_{eq}\exp[-(\Omega_b (1+\Omega_{m}^{-1}(2{\rm
h})^{1/2}-0.06))]\,  \nonumber 
\end{equation} 
works reasonably well, to about 3\% over the region most
relevant to LSS; replacing $\Gamma$ by $\Gamma_{\rm eff} = \Gamma
+(n_s-1)/2$ takes into account the main effect of spectral tilt over
the LSS wavenumber band (\npbh95). For low redshift clusters, the abundances
determine a combination that is roughly $\sigma_8 \Omega_m^{0.56}$
(with the degeneracy among the combination broken with high redshift
cluster information). Weak lensing determines a similar combination.

With the wealth of data emerging from the Sloan Digital Sky Survey (SDSS) and
the 2dF Redshift Survey (2dFRS), shape is a very powerful probe. However, the
biasing of the galaxy distribution with respect to the mass becomes an issue
if it is scale dependent, inviting caution -- and for our purposes a
weakened prior over what the data formally show. In the future, weak
lensing should allow shape and amplitude to be simultaneously
constrained without biasing uncertainties.

To be explicit, our prior for $\ln \sigma_8^2$ is of form $\sigma_8
\Omega_m^{0.56}$=$0.47^{+0.02,+0.11}_{-0.02,-0.08}$, distributed as a
Gaussian (first error) smeared by a uniform (tophat) distribution
(second error). This straddles most of the values determined from weak
lensing and many of those estimated from cluster abundances (shown in
figure~\ref{fig:sig8evoln}). We have also used a prior shifted downward
by 15\% to accommodate the lower values quoted for clusters in the
literature. Our prior for the shape parameter is $\Gamma_{\rm
eff}$=$0.21^{+0.03,+0.08}_{-0.03,-0.08}$, which encompasses the recent
SDSS and 2dFRS results as well as results from the Automated Plate
Measuring (APM) angular survey and earlier redshift surveys. Fully
embracing the 2dFRS galaxy power spectrum with a linear bias model to
relate it to the total density power gives a stronger shape
constraint, $\Gamma_{\rm eff}$=$0.21^{+0.03,+0}_{-0.03,-0}$. Although
some SDSS estimates are quite consistent, \eg $\Gamma_{\rm
eff}$=$0.19^{+0.04,+0}_{-0.04,-0}$, different analysis methods and
different datasets give wider ranges and the estimates do not
incorporate possible complexities in the bias model. Thus, we have
adopted a weak-LSS as opposed to a strong-LSS prior. Since $\Gamma$
$\propto \omega_m /h$, with the improved CMB estimations of $\omega_m$
that arose in the \janzerotwo\  (and later) datasets, the shape constraint
now has some similarity to an $h$ prior (\S~\ref{sec:params}).

Values of $\Gamma_{eq}$, $\Gamma$ and $\sigma_8$ as estimated from the
CMB data are given in table~\ref{tab:exptparams}. They are basically
compatible with the LSS priors. One can get $\Gamma_{eff}$ from the
$\Gamma$ and $n_s$ results in the table.

One of the most exciting results from WMAP was the TE
cross-correlation of the E-mode of polarization and the total intensity (T) at low $\ell$,
interpreted as evidence for a $\tau_C = 0.16 \pm 0.04$ detection,
determined with a `model independent' method by Kogut \et\ (2003).  The
detection is not nearly as strong when ensemble-averaged over model
space for the weak prior, as described in \S~\ref{sec:params}. The TE
result is explicitly included in the Cosmomc treatment, but only TT is
included in the ${\cal C}_\ell$ database used here. To incorporate the
detection, we have constructed a $\tau_C$ prior, chosen to be broader
than a $0.16 \pm 0.04$ Gaussian, $\tau_C =
0.16^{+0.04,+0.06}_{+0.04,+0.06}$ in terms of Gaussian and top-hat
errors. The MCMC results for $\tau_C$ we obtain when other parameters
are marginalized is broader still, on both sides, and so we compare
parameter estimates with and without this prior and find for most it
makes little difference It does have an effect on marginalized
amplitude determinations, in particular skewing somewhat the
$\sigma_8$ distribution to higher values.

Sometimes there is `tension' between the parameters estimated from
CMB-only and those including non-CMB priors. This is extremely important to flag,
since poor distribution overlap leads to smaller combined errors.

\subsection{Degeneracy breaking \& parameter eigenmodes} \label{sec:pareigen}

One is tempted to open up parameter space to a much larger set. There
was a good reason for limiting the number in the pre-WMAP days: the
spectra may not change much as the parameters vary, manifested by
near-degeneracies among them. It is useful to disentangle the
degeneracies by making linear combinations which diagonalize the error
correlation matrix $\avrg{\Delta y_a \Delta y_{a^\prime}}$, where $\Delta y_a
\equiv y_a -\avrg{y_a}$ and the averages are over the
probability-weighted ensemble of models.  These `parameter eigenmodes' 
(\npbh95; \npdegeneracies; \nplange00; \npGoldstein02) 
\begin{equation} 
\xi_\alpha = \sum_a
{\cal R}_{\alpha a} \Delta y_a \ \ {\rm  obey} \ \avrg{\xi_\alpha \xi_\beta} =
\delta_{\alpha \beta} \sigma_\alpha^2\, .  \nonumber 
\end{equation} 
The error on the eigenmode
$\xi_\alpha$, $\sigma_\alpha$, is determined by the data and the
priors. (Instead of $\Delta \omega_b$ and $\Delta \omega_{cdm}$, we
use $\Delta \omega_b /\omega_b$ and $\Delta \omega_{cdm}
/\omega_{cdm}$ in the combinations so their errors are relative and
quantitatively meaningful relative to the other variables.)

Until the WMAP data, only four combinations of our seven could be determined
within $\pm 0.1$ accuracy with the CMB (five with CMB+LSS), but with
WMAP precision, for the \marchzerothree\  data, five can be determined (six with
CMB+LSS), and two are determined to better than $\pm 0.01$. Parameter
eigenmodes arising from the current data are discussed further in
\S~\ref{sec:params}\ref{sec:eigen}.  

Thus,  WMAP precision gives us licence to open up the parameter space
more.  Here we only do this to a limited extent, by restricting
ourselves to flat universes and replacing $\Omega_{tot}$ by $w_Q$ or
by $dn_s (k_n) /d\ln k$.

Both MCMC and fixed-grid approaches can have difficulty when the
eigenmodes are precisely determined. Using variables which are
nonlinear combinations of the $\{y_a\}$ motivated by the eigenmodes
can aid with this, \eg one characterizing the peak/dip pattern (the
sound-crossing scale) and one the amplitude-$\tau_C$-$n_s$
near-degeneracy. Both degeneracies were exploited in limiting our
${\cal C}_\ell$ database storage requirements.

 Expressing the LSS prior in terms of $\Gamma + (n_s-1)/2$ and $\ln
\sigma_8^2 \Omega_m^{1.12}$ only (\npBond02; \nplange00; \npbj98) is similar
in spirit to keeping only the best determined `eigenmode' from the
redshift surveys and from the lensing or cluster surveys. However, the
same mechanism that gives the acoustic peaks in ${\cal C}_\ell$ leads
to oscillations in the density power for large $\omega_b / \omega_m$;
\ie further eigenstructure that would be revealed with high precision
shape data. Similarly extra variables such as $\omega_{hdm}$ also lead
to more eigenstructure. Higher redshift observations also break the
$\ln \sigma_8^2 \Omega_m^{1.12}$ near-degeneracy.

\subsection{CMB pillars} \label{sec:pillars}

There were `seven pillars' of the inflation paradigm that we were
looking for in the CMB probe: 

(1) the effects of a large scale
gravitational potential at low multipoles;

(2) the pattern of acoustic
peaks and dips;

(3) damping; 

(4) Gaussianity (maximal randomness for a
given power spectrum) of the primary anisotropies;

(5) secondary anisotropies associated with nonlinear phenomena, due to
the SZ thermal and kinetic effects, inhomogeneous
reionization, weak lensing, \etc; 

(6) polarization, which must be
there at the \ca\  $10\%$ level, along with a specific cross-correlation
with the total intensity; 

 (7) anisotropies and the associated
polarization induced by gravity wave quantum noise.

At least five, and possibly six, of these have been seen. We have
known about pillar 1 since COBE and FIRS, and found pillars 2 and 3 in
the past few years, as discussed in
\S~\ref{sec:analysis}\ref{sec:LsLDLpkphenom},\ref{sec:LsLDLpk}.

Most, but not all, inflation models predict Gaussianity of the primary
CMB fluctuations (pillar 4). This has been demonstrated to varying
degrees with COBE, Maxima, Boomerang, the Cosmic Background Imager
(CBI) and now with WMAP data. All secondary anisotropies and Galactic
foregrounds will be non-Gaussian, so care must be taken in
interpreting the inevitable deviations from Gaussianity.

The CBI, the Arcminute Cosmology Bolometer Array Receiver (\Acbar) and
the Berkeley Illinois Maryland Array (BIMA) may have seen evidence for
the thermal SZ effect, an aspect of pillar 5 (see
\S\S~\ref{sec:cmbdata}\ref{sec:sec} and \ref{sec:params}\ref{sec:margparam}).

Polarization (pillar 6, see \S~\ref{sec:cmbdata}\ref{sec:pol}), has been
convincingly demonstrated. First there was the broadband detection by
DASI of EE polarization and its TE cross-correlation with total
intensity, at levels consistent with inflation models. Then WMAP
unveiled the TE cross-correlation spectrum to $\ell \sim 400$. The
enhancement at $\ell \lta 20$ is the evidence for $\tau_C =0.16 \pm
0.04$ and an associated redshift of reionization $z_{reh} \sim 15$. The WMAP
anticorrelation in TE observed at $\ell \sim 100$ is interpreted as
proof that the dominant component of the perturbations giving rise to
this effect is adiabatic.

Pillar 7 is an extreme experimental challenge, and some inflation
models have gravity wave induced anisotropies too small for  
them ever be detected (see \S~\ref{sec:cmbdata}\ref{sec:pol}).

\section{The CMB experiments}
\label{sec:cmbdata}

\subsection{January 2000} \label{sec:jan00}

We were dealing with upper limits on anisotropies until the
1990s. These limits were useful in ruling out broad ranges of
theoretical possibilities. Then the familiar $2 \le \ell \lta 20$
multipoles at the $30 \mu K$ level were revealed by COBE. This was
followed in the years up to April 1999 by detections, and a few upper
limits (ULs), at higher $\ell$ in 19 other ground-based (gb) or
balloon-borne (bb) experiments. Some predated the 1992 COBE
announcementin design and even in data delivery. We have the
intermediate angle SP91 (gb), the large angle FIRS (bb), both with
strong hints of detection before COBE, and then, post-COBE, more Tenerife
(gb), MAX (bb), MSAM (bb), white-dish (gb, UL), argo (bb), SP94 (gb),
SK93-95 (gb), Python (gb), BAM (bb), CAT (gb), OVRO-22 (gb), SuZIE
(gb, UL), QMAP (bb), VIPER (gb) and Python V (gb).  Most had many
fewer resolution elements than the 600 or so for COBE. One exception
was SK95, which forged new ground compressing raw timestreams in
software on to spatially extended `pixels'. This heterogeneous
dataset up to that time showed evidence for a peak (\npbjkrad00),
although it was not well localized. Improved first-peak localization
occurred in summer 1999 with the Chile-based Toco experiment (\nptoco99)
and in November 1999 with the North American balloon test-flight of
Boomerang (\npmauskopf99). Collectively we denote the results of all
of these experiments as the \janzerozero\  data. It pointed to
$\Omega_{tot}\sim 1$, but with broad errors
(figure~\ref{fig:2Djan000203mar03}).

\vskip 10pt \noindent
{\it Technical notes} \vskip 10pt 

Boom-NA, Toco, QMAP, SK95, MSAM, SP94 and SP91 include
quoted calibration errors; 13 bandpower detections from other
experiments do not explicitly include them,  but are often
incorporated in the quoted error bars. The three bandpower upper limits
use the `equal variance approximation' to the likelihood
(\npbjkrad00), as opposed to the offset lognormal which is used for the
rest. The low quadrupole is not included in the DMR bandpowers.

\subsection{January 2002: Boomerang, Maxima and DASI} \label{sec:jan02}

In April 2000 dramatic results to $\ell \sim 600$ from Boomerang, the
first CMB long duration balloon flight which circled Antarctica for
10.6 days in December 1998, were announced (\npdeBnature00; \nplange00). This
was quickly followed in May 2000 by results from the night flight of
Maxima (\npmaxhanany00). Boomerang's best resolution was $10.7^\prime
\pm 1.4^\prime$, about 40 times better than that of COBE, with tens of
thousands of resolution elements. (The corresponding Gaussian beam
filtering scale in multipole space is $\sim 800$.) Maxima had a
similar resolution but covered an order of magnitude less sky.  In
April 2001, the Boomerang analysis was improved and much more of the
data were included, delivering information on the spectrum up to $\ell
\sim 1000$ (\npnett01). Maxima also increased its $\ell$ range
(\npmaxlee01).

Boomerang had six bolometers at 150 GHz and 10 other bolometers at 90,
220 and 400 GHz. It mapped 1800 square degrees. The April 2000 analysis
used only one channel and 440 $deg^2$.  The April 2001 analysis used 4
channels at 150 GHz and 800 $deg^2$, $1.8\% $ of the sky. These are the
data used in the \janzerotwo\  and \junzerotwo\  sets (\npnett01).  In December 2002,
the final Boomerang-98 analysis was given (\npRuhl02), encompassing an
area 1200 $deg^2$, $f_{sky} = 2.9\%$, using $3.5^\prime$ pixels and the
150 GHz channels. This is part of the \janzerothree\  set. 

Boomerang had a successful second long-duration balloon flight in
January 2003, taking data for 13 days.  The forecasts are for a well
determined TT power spectrum even in the $\ell > 1000$ regime,  because
the beam uncertainty is smaller than for the first flight. The
150 GHz detectors were polarization sensitive bolometers of the sort
that will be used on Planck, and polarizing grids were used with the
240 and 340 GHz bolometers: good multi-band EE and TE spectra should
emerge as well.

 The South-Pole-based DASI (the Degree Angular Scale Interferometer)
has 13 dishes of diameter 0.2 m and uses high-electron-mobility
transistors (HEMTs) spanning $26$--$36$GHz. An interferometer baseline
directly translates into a Fourier mode on the sky. The dish spacing
and operating frequency dictate the $\ell$ range. In DASI's case, the
range covered is $125 \lta \ell \lta 900$. The total area covered was
288 $deg^2$ consisting of 32 independent maps of size $3.4^\circ$, the
field-of-view.  In April 2001, DASI unveiled a spectrum (\npDASI01)
similar to that reported by Boomerang at the same time. The two
results together reinforced each other and lent considerable
confidence to the emerging (\janzerotwo) ${ \cal C}_\ell$ spectrum in the
$\ell <1000$ regime.

\vskip 8pt \noindent
{\it Technical notes} \vskip 6pt 

 For Maxima, 13 bandpowers, covering $36 \le \ell \le
1235$, were used, with a 4\% calibration uncertainty and a 5\% beam
uncertainty. For both Boomerang sets, the power spectra use the
`Faster' method on $3.5^\prime$ pixels, the calibration error is
10\% and the beam uncertainty is 13\%. Of the 25 bands, $2$--$21$ were
used in our parameter analyses, covering $26 \le \ell \le 1025$. The
other bands are marginalized. For DASI, the calibration uncertainty is
4\%, there is no beam uncertainty, and all 9 bands are used in our
analyses.

\subsection{June 2002: CBI, VSA and BIMA} \label{sec:jun02}

The CBI (Cosmic Background Imager) is based 17000 feet above sea level
on the Atacama Plateau in Chile. It has 13 dishes, 0.9 m in diameter,
operating in the same HEMT channels as DASI. The instrument measures
78 baselines simultaneously. The larger dishes and baselines imply
higher resolution, to $\ell $ of 3500, a huge increase over Boomerang,
Maxima and DASI. Only the analyses of data from the year 2000
observing campaign were reported in May 2002, covering three deep
fields, of diameter \ca\ $0.75^\circ$ (\npMason02b), and three mosaic
regions, each of size \ca\ $13$ $deg^2$ (\npPearson02). Mosaics lace
together multiple interferometer fields-of-view (FOVs) which overlap,
allowing greatly improved resolution in $\Delta \ell$ over what many
independent FOVs can give.

CBI data from 2001 roughly doubled the total amount available and
increased the area covered. The combined two-year data are for three mosaic
fields covering about 80 $deg^2$, including the two deep fields which
overlap, and for one disjoint deep field. Preliminary results for the
CBI 2000+2001 data were reported at the Royal Society
meeting (January 2003). A full analysis is given in \Readhead03.

The Very Small Array (VSA) in Tenerife, also an interferometer
operating at 30 GHz, covered the $\ell$ range of DASI, and confirmed
the spectrum emerging from the Boomerang, Maxima and DASI data in that
region in May 2002 (\npVSA02). It is part of the \junzerotwo\  set.  The VSA
observed at longer baselines to increase its $\ell$ range and
announced results in December 2002 (\npVSAext02). It is included in the
\janzerothree\  set.

In \junzerotwo, the interferometric millimetre array BIMA, operating instead at 30
GHz, announced power at $\ell \sim 6000$ had been found in random
fields (\npBIMA02). This is too far out in $\ell$ to be used for the
analysis of primary anisotropies, but is important for secondary
anisotropies (see \S~\ref{sec:cmbdata}\ref{sec:sec}).

\vskip 8pt \noindent
{\it Technical notes} \vskip 6pt 

 The one-year CBI mosaic and deep field data use a
conservative calibration error of 5\%. For parameters, bands $1$--$12$ 
covering $\ell$ to 1940 are used in the odd binning of the mosaic
power spectrum of Pearson \et\ (2003).  The 10 bands at higher $\ell$
are marginalized.  When the deep field data are used in conjunction
with the mosaic for optimal spectra, all bands are used. The deep data
are not used for cosmic parameter estimation. The two-year data had a
3.3\% calibration error as of \janzerothree, but with WMAP's observation of
Jupiter this could be reduced to 1.3\%, with a 3\% recalibration
downward in amplitude. The \junzerotwo\  VSA results have a calibration
uncertainty of 3.5\%, and have 10 bands up to $\ell$ of 1000. The
\janzerothree\  extended VSA results have the same calibration, and 16 bands
extending to $\ell$ of 1700. For \marchzerothree\  they were recalibrated with the
lower calibration uncertainty, as for CBI.

\subsection{January 2003: \Archeops\ and \Acbar; Extended Boomerang, 
 VSA \& CBI} \label{sec:jan03}

The \janzerothree\  data mix replaces Boomerang, the VSA and CBI by their
extended sky results described in \S~\ref{sec:cmbdata}\ref{sec:jan02},\ref{sec:jun02}.  

\Archeops\ mapped 30\% of the sky at a FWHM resolution of
$15^\prime$ in a 22 hour balloon flight. It used the bolometer
detectors and scanning strategy similar to those that will be employed
for the Planck satellite.  In October 2002, the \Archeops\ team
presented a power spectrum to $\ell =350$ that overlapped with the
COBE/DMR power spectrum at low $\ell$ (\nparcheops02), derived from 12\%
of the sky, one of their six 143 GHz bolometers and one of their six
217 GHz bolometers. 

\Acbar\ was installed in January 2001 on the 2.1 m diameter Viper
telescope at the South Pole. It has 16 bolometers in three frequency
bands centered at 150, 220 and 280 GHz, with a best resolution of 
\ca\ $4^\prime$, allowing excellent coverage to high $\ell$. The
analysis to obtain the power spectrum of Kuo \et\ (2002) used the four
bolometers at 150 GHz available in the first observing season, the eight
available in the second, and covered $\sim 24$ $deg^2$ in two fields
consisting of correlated $\sim 3 \times 1.5$ $deg^2$ patches.  The
cosmological parameters determined with its power spectrum in
conjunction with other data were given in Goldstein \et\
(2002). \Acbar\ is being further upgraded and continues to operate.

In Sept 2002, DASI announced the detection of polarization, and this
is discussed in \S~\ref{sec:cmbdata}\ref{sec:pol} and as Pillar 6.

\vskip 8pt \noindent
{\it Technical notes} \vskip 6pt 

  The \Acbar\ set has a calibration uncertainty of 10\%
and an uncertainty of 3\% on its $\sim 5.5^\prime$ beam at 150
GHz. For the analyses here, bands $2$--$14$ are used, covering $300 \le
\ell \le 3000$, with the first bin marginalized. (The window function
of the first bin is oscillatory and reaches reasonably high
$\ell$. However, for optimal spectra and parameter results in
Goldstein \et\ (2003), the first bin has also been used, and yields
similar results to those obtained without it, as described below.)
\Archeops\ has a calibration uncertainty of 7\% and an uncertainty of
10\% on its $15^\prime$ beam. Sixteen bands covering $15 \le \ell \le 350$
are used.
  
\subsection{March 2003: WMAP} \label{sec:mar03}

In June 2001, NASA launched the five-frequency all-sky HEMT-based WMAP
satellite. The  best resolution of $12.6^\prime$ at 94 GHz 
corresponds to a `Gaussian' beam size of $\ell \approx 640$. The
first-year results unveiled in February 2003 were as spectacular as
forecast, definitive through the second peak. The noisy error bars at
higher $\ell$ will subside as the observing period increases, and the
beam is very well known so that we may expect good spectra out to $\ell
\sim 1000$ with four years of data.

COBE was fundamental to every parameter determination with pre-WMAP
data.  The WMAP verification that the COBE/DMR maps were accurate in
detail was by itself an important step.  Pillar 1's Sachs-Wolfe effect
dominates at low $\ell$.  It includes both the `ordinary' effect
from fluctuations in $\Phi$ on the last scattering surface and the
`integrated' Sachs-Wolfe (ISW) effect, from the change of $\Phi$ with
time so the energy of a photon climbing out of a potential well
differs from that when it dropped in.  Other effects can influence low
$\ell$, the contribution from tensor modes, and more exotic
possibilities involving low mass scalar fields, modified topologies,
radically broken scale invariance, \etc\ For the best-fit $\Lambda$CDM
models, an upturn in ${\cal C}_\ell$ from the ISW is predicted, but
instead a downturn is observed with WMAP, confirming and extending the
puzzles associated with the relatively low quadrupole. It has been our
practice to drop the $\ell=2$ mode in parameter estimation
(marginalize over it), and we continue it here, but recognize it could
be pointing to new physics. The quadrupole has such large cosmic
variance that including it or not does not change parameter
determination in the minimal inflation models much.

The low-$\ell$ DMR data has been used to constrain the size of the
Universe, basically from the scale of its hot and cold spots. If our
manifold was much smaller than the apparent distance to the last
scattering surface, there would not be large-scale spots in DMR maps,
and smaller scale ones would be images of each other, with placement
and details dependent upon specifics of the manifold. The complication
here is that there are many possible manifolds, and each has an
orientation, so it is possible to find universes that are just big
enough to be more probable than the conventional inflation models with
far fewer degrees of freedom in which the large-scale hot and cold
spots are not geometrically correlated as in the topology case. See
\BPS00top for a discussion.  WMAP data should allow more powerful
tests and improve the current constraints on size somewhat.

\vskip 8pt \noindent
{\it Technical notes} \vskip 6pt 

 The WMAP power spectra for our analyses are those
obtained in Hinshaw \et\ (2003) for TT to $\ell = 900$ and in Kogut \et\ 
(2003) for TE to $\ell =512$, using the construction in Verde \et\ (2003) 
for the correlated errors. 

For our ${\cal C}_\ell$ database approach, we include the 0.5\%
calibration uncertainty, use the offset lognormal approximation to the
likelihood surface, and, as for DMR, we marginalize over the WMAP
quadrupole. We have compressed the data to speed up our analyses. We
have used the full 899 individual multipoles of \wmapH\ and \wmapV, 
including correlations out to a (tiny) sideband cutoff. Doing many
large matrix multiplies can be slow, as relaxation to the amplitudes
is done for each of the 8.5 million model elements in the
database. Although these operations could be sped up considerably by approximate
log-likelihood estimates to reject extremely improbable models, and there 
are many of these, we have so far used brute force through all models. 

We have also used compressions of the WMAP data onto fewer bands for
parameter estimations, optimal spectra and calibration estimation. For
example, a 98-band compression of the data preserves all signal and
noise information and side-band correlations in the bands. has $\Delta
\ell$ spacings ranging from $1$ at small $\ell$, to 7 to 5 to 7
through $\ell$=373, 9 through 565, with a gradual increase beyond 600, and
only 4 broad bands beyond 700 (with relative errors above 50\%).  A
best fit shape was used in the compression, but other choices,
including a flat shape, give very similar spectra where WMAP dominates
the data.  In a 49-band compression, the band spacings are essentially
doubled, and in a 19-band compression, the Boomerang $\Delta \ell =50$
spacing was used. Parameter results using the ${\cal C}_\ell$ database
are quite consistent among the 899-band, 98-band and 49-band sets,
with slightly sharper errors with the 899-band set. We do find small
(sub-$1\sigma$) deviations in parameters that are bundled into some
of the exquisitely determined eigenmodes for the 899 set \cf\ the
98 set.

For our MCMC calculations, the WMAP likelihood routines of Verde \et\ 
(2003) are adopted, since these are used in the Cosmomc package
(\npcosmomc). This uses a hybrid of an offset lognormal and a Gaussian
distribution to compute parameters.  The quadrupole is included. The
small calibration error of 0.5\% is not. For ${\cal C}_\ell$
construction, CMBfast was used for the database and CAMB is used in
Cosmomc. Other differences between the database and MCMC treatments of
WMAP are bigger, \eg mimicking the effects of the TE data with a
$\tau_C$ prior. The parameter results for the 98-band case using
MCMC are in good agreement with those using the  database.

\subsection{Beyond March 2003: Planck and targeting polarization} \label{sec:pol}

Many other CMB experiments on the ground and in balloons will happen
before the European Space Agency's (ESA's) Planck satellite is launched in
2007. Planck combines bolometers, many of which will be polarization
sensitive, and HEMTs. Its best resolution of $\sim 5^\prime$ and
detector sensitivity should allow the damping tail and power spectrum
modifications due to secondary effects, including weak lensing, to be
very well determined.  The all-sky component-separated intensity and
polarization maps will be superb for interstellar matter and
extragalactic source research as well as cosmological research. Such
accuracy is needed to open up the cosmic parameter space and search
for anomalies that may signal new physics beyond the minimal inflation
concordance model we are drawn to now.

Polarization is described by a 2$\times$2 tensor on the sky, with the
components related to the four Stokes parameters, $T,Q,U,V$, with $T$
the total intensity (\ie temperature). The polarization dependence of
Compton scattering induces a well-defined polarization signal emerging
from photon decoupling, arising from the quadrupole nature of the
viscosity-induced anisotropic stress tensor. For primary CMB
fluctuations, circular polarization is not there and $V$ vanishes. When
$Q$ and $U$ maps are Fourier transformed and are rotated into a basis
related to the angular wavevector, these give $E$-type (grad) and
$B$-type (curl) `maps'. Scalar perturbations in linear theory do not
generate $B$-type modes, so their absence is a check. Tensor perturbations
generate both, and detection of $B$-modes at low $\ell$ would be a
direct signature of a gravitational wave background (pillar 7).

Given the total ${\cal C}_\ell^{(TT)}$ of figure~\ref{fig:CLoptmar03},
we can forecast the polarization power ${\cal C}_\ell^{(EE)}$ and
cross-correlation power ${\cal C}_\ell^{(TE)}$: the maximum signal is
expected at $\ell \sim 900$, with amplitude $\sim 5 \mu$K over $\ell
\sim 400-1600$.

The great race to first detect CMB $E$-mode polarization was won by
DASI (\npDASIpol02), with 271 days of polarization data on rwo deep fields
($3.4^\circ$ FOV) showing a $5\sigma$ detection with a value
$0.8 \pm 0.3$ of the forecasted amplitude from $T$ for inflation-based
models. The cross-correlation of the polarization with the total
anisotropy had an amplitude $0.9 \pm 0.4$ of the forecast. These
detections used a broadband shape covering the $\ell$ range $\sim
250-750$ derived from the theoretical forecasts.  The powerful
cosmological implications of the remarkable WMAP TE bandpowers have 
already been discussed.

Forecasts indicate solid EE power spectrum determinations are likely
soon from the ongoing CBI polarization observations and Boomerang's
January 2003 flight with polarization-sensitive bolometers. Both are
optimally sensitive to the $\ell \sim 900$ region where the EE power
is expected to peak.  MAXIMA will fly again as the
polarization-targeting MAXIPOL. Other EE experiments, operating or
planned, include AMiBA, CAPMAP/PIQUE, COMPASS, CUPMAP, POLAR,
Polarbear, Polatron, QUEST and Sport/BaRSport, among others. We are also
awaiting WMAP's EE results.

Although the strength of the $B$-mode induced by gravity waves is
model specific, the amplitude is expected to be quite small even
at low $\ell$. Nonetheless there are experiments such as BICEP being
planned to go after ${\cal C}_\ell^{(BB)}$ in these low $\ell$
ranges. Planck could also make such a detection. A nice figure
summarizing EE and BB bandpower forecasts for various experiments is
given in Hivon \& Kamionkowski (2002).  The promise is sufficiently
exciting that a CMBPol satellite is being contemplated by NASA as the
next step in space for the CMB after Planck.

\subsection{Beyond March 2003: targeting  secondary anisotropies} \label{sec:sec}

 Spectral distortions from the CMB black body must exist as a result
of nonlinear processes and will have associated anisotropies.  The
spectrally well-defined SZ distortion associated with Compton
upscattering of CMB photons from hot gas has not been observed in the
spectrum. The COBE/FIRAS 95\% upper limit of $ 6 \times 10^{-5}$ of
the energy in the CMB is compatible with values up to around $10^{-5}$
expected from the cosmic web of clusters, groups and filaments in the
inflation-based class of models considered here, and places strong
constraints on the allowed amount of earlier energy injection, \eg
ruling out mostly hydrodynamic models of LSS.

The SZ effect has been well observed at high resolution with very high
signal-to-noise along lines-of-sight through a large number of
clusters now, by single dishes, the OVRO and BIMA millimetre arrays, and the
Ryle interferometer. This tells us, among other things, that the CMB
comes from further away than redshift $z \sim 1$ -- if we had any
residual doubt. The SZ effect in random fields may be responsible for
the power at $\ell > 2000$ seen in the CBI deep data
(figure~\ref{fig:2Djan000203mar03}) (\npMason02b; \npBond02), in the BIMA
data at $\ell \sim 6000$ (\npBIMA02), and possibly in the \Acbar\ data
(\npGoldstein02). Multi-frequency observations to differentiate the
signal from the CMB primary and radio source contributions will be
needed to show this.

A number of planned HEMT-based interferometers are being built with
this ambient effect as a target: CARMA (OVRO+BIMA together), the SZA
(from Chicago, and to be incorporated in CARMA), AMI (based in
Britain, including the Ryle telescope) and AMiBA (from
Taiwan). Bolometer-based experiments will also be used to probe the SZ
effect, including: the CSO (Caltech Sub-Millimetre Observatory, a 10 m
diameter dish) with BOLOCAM on Mauna Kea; the LMT (Large Millimetre
Telescope, with a 50 m diameter dish) in Mexico; APEX, a 12 m diameter
German single dish based in Atacama; Kobyama, a 10 m diameter Japanese
single dish; and the 100 m Green Bank telescope. Large bolometer
arrays with thousands of elements and resolution below $2^\prime$ are
also under development: the South Pole Telescope (SPT, Chicago) and
the Atacama Cosmology Telescope (ACT, Princeton).

The kinetic SZ effect due to the motion of clusters in the
low-to-moderate-redshift cosmic web, or just clumpy ionized gas in
the high redshift cosmic web, has the same spectral signature as
primary anisotropies, and so is harder to disentangle.

The (non-Galactic) distortions from the black body that have been
detected in the COBE FIRAS and DIRBE data are associated with
starbursting galaxies due to stellar and accretion-disk radiation
downshifted into the infrared by dust then redshifted into the
sub-millimetre. This background has energy about twice the total of
that in optical light, about a tenth of a percent of that in the CMB.
About $50\%$ of this sub-millimetre background has been identified with
sources found with the SCUBA bolometer array on the JCMT. Anisotropies
from dust emission from these high redshift galaxies are being
targeted by the JCMT (with a SCUBA-2 very large array in development),
the OVRO millimetre interferometer, the CSO, the SMA (Sub-Millimetre Array) on
Mauna Kea, the LMT, the ambitious US/ESO ALMA millimetre array in
Chile, the LDB BLAST, and ESA's Herschel satellite.

\section{CMB analysis and phenomenology} \label{sec:analysis} 

\subsection{CMB pipelines: from timestreams or visibilities to bandpowers} \label{sec:pipe} 

Analysing CMB experiments involves a pipeline that takes the raw
detector timestreams or visibilities from correlators for
interferometers, flags and cleans them, and usually generates maps, from
which bandpowers and higher order statistics are derived, ideally
after separating component signals by using their differing frequency
and spatial dependences. The step from bandpowers to cosmic parameters
described in \S~\ref{sec:GUSparam}\ref{sec:bandtoparams} may be the
goal, but it is not where most of the time is spent. Indeed parameter
determination is used as a diagnostic along with everything else as
the CMB teams struggle to understand in detail their experimental
results. Every new round of data generates a fresh look at pipelines,
and often new faster algorithms for proceeding. Recent pipelines are
described in \eg \bc01, \nett01, Hivon \et\ (2002) and \Ruhl02\ for
single-dish bolometers, in \eg \Myers02\ for interferometry, and in
\wmapB, \wmapH\ and \wmapV\ for WMAP.

For single-dish experiments, the timestreams are turned into spatial
maps for each frequency: an average temperature in each pixel and a
pixel-pixel noise correlation matrix from which the bandpowers, noise
in the bandpowers and band-band error matrices are derived. The first
step is to extract the sky signal from the noise, using the only
information we have, the pointing matrix mapping a bit in time onto a
pixel position on the sky. In the analysis of Boomerang, and
subsequent work including for WMAP, powerful use of Monte Carlo
simulations was made to evaluate the power spectrum and other
statistical indicators in maps with many more pixels than was possible
with the conventional matrix methods described in \bjkquad98\ and
Borrill (1999). 

For interferometer experiments, the basic data are visibilities as a
function of baseline and frequency, with contributions from random
detector noise as well as from the sky signals. A baseline is a direct
probe of a given angular wavenumber vector on the sky, and hence suggests
that we should make `generalized pixel' maps in `momentum space' (\ie
Fourier transform space) rather than in position space, as for
Boomerang. For CBI, the $> {\cal O}(10^5)$ visibility measurements of
each field were `optimally' compressed into a $ < {\cal O}(10^4)$
coarse-grained lattice in momentum space from which the power spectrum
was  calculated using matrix methods (\npMyers02).

The important step of separating multi-frequency timestream data into
the physical components on the sky is fundamental, still under active
development, and will remain so for a long time, as our precision
increases. The sources are the primary CMB, the thermal and kinetic SZ
effects, the dust, synchrotron and bremsstrahlung Galactic signals,
the extragalactic radio and sub-millimetre sources, possibly spinning
dust, and of course the sources we have not thought of yet. An example
is the treatment of point sources: at the high resolution of CBI and
its 30 GHz frequency, the contribution from extragalactic radio
sources is significant, so known point sources were projected out of
the visibilities by using a large number of constraint-template
matrices which marginalize over all affected modes, using positions
from the (1.4 GHz) National Radio Observatory Very Large Array Sky
Survey (NVSS) catalog (\npMason02b). Even though the WMAP resolution is
substantially lower than that of CBI, WMAP cut out some 700 sources in
its analysis.  Boomerang cut out three quasars.

For extended sources, the most expedient method is to just cut out
problematic areas, \eg with Galactic latitude cuts to remove emission
from the plane, as was done for DMR, WMAP and Boomerang. Extensive use
is also made of templates that come from Galactic observations, \eg
the IRAS/DIRBE maps of sub-millimetre emission, HI maps, \etc, and from the
experiments themselves, with the highest and lowest frequency maps
often being completely dominated by foregrounds (as for the
dust-dominated 400 GHz Boomerang map).  Having errors on primary
signal maps that reflect the residual contamination after separation
is a priority to ensure that the precision is unbiased as well as
high.

The extension of these methods to polarization is under active
development. For CBI, the polarization pipeline using generalizations
of the \Myers02 techniques has been implemented.  For Boomerang, the
basic tools being developed include a generalization of the
signal-noise separator used in Boomerang-98 (Prunet \et\ 2001; Dore \et\ 
2002), and a variety of power spectrum estimators such as Spice
(correlation function techniques (Chon \et\  2003)), Faster (approximate
$\ell$-space techniques (\npnett01; \npRuhl02)) and Madcap (full matrix
methods (Borrill 1999)).

\subsection{Calibrating by power spectra} \label{sec:calib}

The maximum-likelihood values of the calibration and beam parameters
and the errors on them that are determined in the calculation of the
optimal power spectra of \S~\ref{sec:GUSparam}\ref{sec:GUS} are
themselves very useful: for the power spectra to be consistent with
one derived from an underlying Gaussian model for the anisotropies,
the experimentally quoted uncertainties are often reduced, as shown in
table~\ref{tab:calib}.  Even without the excellent overall amplitude
accuracy to 0.5\% of WMAP, self-calibration among the pre-WMAP
experiments was giving some of the same factors, which is why the
grand unified spectra look so good relative to WMAP. Overall
figure~\ref{fig:CLoptevoln} shows an amazing concordance of data.
Direct calibration using source observations or maps on equivalent
regions is preferable. This has been done for Boomerang with WMAP maps
by Hivon ($\approx 0.95 \pm 0.02$) and for CBI with WMAP observations
of Jupiter, the CBI calibrating source. It is remarkable how close
these recalibrations are to those given here using global power
spectrum analyses.
 
The CBI one-year data started with a $\pm 0.05$ calibration
error. Running the optimal spectrum with WMAP, CBI is predicted to
have its calibration lowered by $0.970 \pm .037$.  For the two-year CBI
data, we begin with the \janzerothree\  uncertainty of $\pm 0.033$ and get
$0.970 \pm .022$.  The Jupiter calibration with WMAP resulted in a 3\%
downward calibration with a 1.3\% calibration error (\npReadhead03).
The same calibration adjustment and uncertainty apply to the VSA,
which also used Jupiter.

\vskip 8pt \noindent
{\it Technical notes} \vskip 6pt 

 The WMAP power spectrum used for this calibration
and beam analysis was the 49-band compression of the $\ell \le 900$
spectrum and the optimal compression was onto the 18 bands of \junzerotwo\ 
in figure~\ref{fig:CLoptevoln}, but the results are very robust: other
compressions for either WMAP or the optimal spectra, varying the
${\cal C}_\ell$ shape of figure~\ref{fig:CLoptevoln} used to get the
bandpowers, and adding more experiments in the calibrating mix make
little difference.

\begin{table}
\caption{Calibration parameters that the temperature anisotropies
should be multiplied by, as determined in the computation of optimal
spectra. {\small\rm (Before WMAP, calibrations by this method were made relative
to each other, with the interferometry data (INT) dominating because
of smaller uncertainties: 4\% for DASI; 5\% for CBI in the \junzerotwo\  data,
3.3\% in \janzerothree, and 1.3\% in \marchzerothree\  (\npReadhead03); 3.5\% for VSA for \junzerotwo\ 
and \janzerothree, and 1.3\% in \marchzerothree.  WMAP gives 0.5\%. INTc denotes the
recalibrated CBI and VSA, along with DASI. The Boomerang calibrations
shown are for the 2.9\% sky-cut Ruhl \et\ (2003) `Faster' power
spectrum, in conjunction with INT and/or WMAP. The values are quite
robust as other experiments are added. (Similar values are obtained if
we use the 1.8\% Netterfield \et\ (2002) cut rather than the 2.9\%:
$0.953 \pm 0.037$, $0.938 \pm 0.036$, $0.949 \pm 0.021$ and $0.942 \pm
0.020$ for INT, INTc, WMAP and WMAP+INTc.) \Acbar\ was done with
Boomerang included and the 2--14 bins. (The multipliers are lower
when the first bin is included, $0.94 \pm 0.05$, $0.92 \pm 0.05$,
$0.89 \pm 0.047$, $0.914 \pm 0.042$.) \Archeops\ was also done with
Boomerang included for INT and INTc. The results are the same
whether Boomerang is recalibrated or not. Beam adjustments are also
computed by the method. For example, the Boomerang beam is compatible with no
change, but substantially reduced errors: the prior is $1.00 \pm 0.13$
and we determine $1.02 \pm 0.03$ with INT, $0.984 \pm 0.026$ with WMAP
and $1.00 \pm 0.022$ with WMAP+INTc.  When cosmological parameters are
determined, adjustments such as these are done for each experiment to
maximize the entropy of each model in turn, but the uncertainties are
then marginalized. The optimal bandpowers also have the calibrations
marginalized.)}  }
\label{tab:calib}
\begin{center}
\begin{tabular}{|l|l|l|l|}
\hline
Calib   & Boom & \Acbar & \Archeops   \\
\hline
prior& $1.00 \pm 0.10$ & $1.00 \pm 0.1$ & $1.00 \pm 0.07$   \\ 
INT &  $0.975 \pm 0.035$ & $0.96 \pm 0.05$ & $1.06 \pm 0.034$   \\
INTc &  $0.961 \pm 0.036$ & $0.947 \pm 0.044$ & $1.05 \pm 0.039$    \\
WMAP & $ 0.963 \pm 0.017$ &  $0.912 \pm 0.049$ &  $1.046 \pm 0.019$   \\ 
WMAP+INTc &  $0.959 \pm 0.017$ &  $0.938 \pm 0.044$ &  $1.046 \pm 0.019$   \\
\hline
\end{tabular}
\end{center}
\end{table}

\subsection{The phenomenology of peaks and dips } \label{sec:LsLDLpkphenom}

The emerging structure in the evolving power spectra of
figure~\ref{fig:CLoptevoln} has stimulated a number of
`model-independent'  approaches to determining the statistical
significance of peaks and dips. Here we apply a simple procedure we
have used on Boomerang and CBI. We model the local band-power profile
as the quadratic form 
\begin{equation} 
{\cal C}_{\rm b} = {\cal C}_{pk} +\kappa_{ pk}
\langle ({\ell} -{\ell}_{ pk})^2\rangle_{\rm b} /2\, ,   \nonumber 
\end{equation} 
where $b=1, .., N$ runs over the number of bands $N$ we include as we
slide over the data, and $\langle \rangle_{\rm b} $ denotes
band-average.  The position ${\ell}_{pk}$, amplitude ${\cal C}_{pk}$,
and curvature $\kappa_{pk}$, are treated as `internal' parameters,
just as calibration and beam uncertainties are: they relax to their
maximum-likelihood values and errors are estimated from the curvature
matrix of the likelihood function about this maximum.  A numerical
indication of the significance of a detection is the number of sigma
the peak curvature differs from zero,
$|\kappa_{pk.m}|/\sigma_{\kappa_{pk}}$. To detect a peak, we require
that this significance exceeds unity, and also that $\ell_{pk,m}$ lies
within the range of multipoles covered by the bands.  The results of
applying this algorithm to the evolving grand unified spectra and to
WMAP are shown in table~\ref{tab:Lpkslider}.  No attempt was made to
optimize these detections by exploring band positioning and spacings.

The first peak was hinted at in the April 1999 data, was seen in the
\janzerozero\  data when Toco and Boomerang-NA were added, and localization
improved by \janzerotwo\  with Boomerang, Maxima, and DASI, and then with
\Archeops\ by \janzerothree, and now to very high accuracy by WMAP.
Boomerang+DASI (2001) detected the second and third peaks, first and
second dips, and CBI (2002) detected the second, third, and fourth
peaks and the third and fourth dips, albeit with some at low
significance.

\vskip 8pt \noindent
{\it Technical notes} \vskip 6pt  

  The slider bandwidth $N$ we choose depends upon the
band-spacing. For the unified spectra of the evolving data, we
compressed onto the $\Delta \ell$=50 bins used for Boomerang up to
900, then onto a `CBI odd' binning used for the two-year data,
although the CBI bands used were for the `CBI even' binning: the odd
binning is more similar to the \Acbar\ spacing. For the \junzerotwo\  data,
CBI one-year $\Delta \ell =140$ odd bins were used. Sliders of width
$N = 3,4, 5$ were tried, $N=3$ being more appropriate for high $\ell$ values
and $N=4$ being more appropriate for lower values, to keep the the
$\ell$-range probed limited to a few hundred.  WMAP data were first
compressed onto this Boomerang binning before the optimal spectra
were constructed for the \marchzerothree\  data, but results are relatively
insensitive to this and to band locations and widths. For \Archeops,
the $\Delta \ell$ are smaller,  so larger $N$ gives the most robust
detections.

\begin{table}
\caption{Peak/dip locations $\ell_{pk/dip,j}$ and heights ${\cal
C}_{pk/dip,j}$ (in $\mu K^2$) from maximum-likelihood analysis of
data within $N$ bands as the bandwidth slides over the data.  {\small\rm (Results
for a four-band-width slider acting on the compressed grand unified
spectra of the various epochs are shown.  Values of the curvature
$|\kappa_{pk/dip, m}|$ in units of $\sigma_{\kappa_{pk}}$ are given
beside the $\ell_{pk/dip,j}$. A detection requires that
this exceeds unity and $\ell_{pk/dip,j}$ lies within the bands
probed. If a peak/dip detected with a three-band slider is more
significant than for the four-band slider, it is shown in brackets. When we do
WMAP-only with this crude Boomerang binning we get the results in the
second column that are in excellent accord with the Gaussian-fit
results of \wmapP: ($220 \pm 0.8$, $5373 \pm 337$), ($546 \pm 10$,
$2381 \pm 83$) and ($411.7 \pm 3.5$, $1707 \pm 43$) for the first and
second peak and the first dip.  Other individual experiments we have
applied this procedure to include Boomerang, CBI and \Archeops.  The
numbers in the table also accord well with those we have obtained for
Boomerang (\npdeBpkdip01; \npRuhl02) and CBI (\npPearson02; \npReadhead03)
data alone; \eg for Boomerang-only with the Faster power spectrum,
recalibrated as in table~\ref{tab:calib}, we get ($221 \pm 7$, 3.5,
$5371 \pm 337$), ($528 \pm 15$, 1.5, $2183 \pm 136$), ($820 \pm 15$,
1.5, $2000 \pm 215$) for peaks and ($412 \pm 7$, 5.2, $1706 \pm 105$),
($699 \pm 21$, 1.5, $1708\pm 139$) for dips.  \archeops02\ report the
\Archeops\ first peak location to be at $220 \pm 6$; with our method,
all sliders of width $N\ge 4$ detect the first peak accurately; \eg $N=9$
probing $\ell$ values from 110 to 350 give ($221 \pm 6$, 5.7, $5296 \pm
360$). The recalibration of \Archeops\ of table~\ref{tab:calib} has been
applied to ${\cal C}_{pk,1}$, bringing it into good accord with the
WMAP and \marchzerothree\  numbers.) }}
\label{tab:Lpkslider}
\begin{center}
\begin{tabular}{|l|lllll|}
\hline
   & WMAP & Jan02 & Jun02 & Jan03 &  Mar03  \\
\hline
$\ell_{pk,1}$ & $221^{+1}_{-1}$, 20   & $219^{+6}_{-6}$, 3.5      & $218^{+6}_{-6}$, 3.7        & $220^{+5}_{-5}$, 4.5 & $221^{+1}_{-1}$, 20 \\
$\ell_{pk,2}$ & $537^{+12}_{-12}$, 2.9     & $538^{+17}_{-17}$, 1.6       & $536^{+14}_{-14}$, 1.8      & $535^{+12}_{-12}$, 2.0 & $535^{+8}_{-8}$, 3.8    \\
$\ell_{pk,3}$ &  &  ($805^{+10}_{-10}$, 1.5)         & $835^{+16}_{-16}$, 2.0      & $826^{+11}_{-11}$, 2.2  & $823^{+12}_{-12}$, 2.0  \\
$\ell_{pk,4}$ &  &                                 &       & ($1141^{+13}_{-13}$, 2.2) & ($1141^{+13}_{-13}$, 2.2)  \\
\hline
${\cal C}_{pk,1}$ & $5386^{+58}_{-58}$ & $5393^{+443}_{-443}$           & $5501^{+392}_{-392}$  & $5414^{+325}_{-325}$ & $5388^{+57}_{-57}$ \\
${\cal C}_{pk,2}$ & $2326^{+85}_{-85}$ & $2351^{+196}_{-196}$           & $2350^{+164}_{-164}$ & $2364^{+140}_{-140}$ & $2313^{+67}_{-67}$  \\
${\cal C}_{pk,3}$ &  &  $2565^{+400}_{-400}$   & $2296^{+256}_{-256}$ & $2490^{+231}_{-231}$ & $2272^{+175}_{-175}$   \\
${\cal C}_{pk,4}$ &   &                               &                      & $1279^{+241}_{-241}$ & $1219^{+227}_{-227}$  \\
\hline
$\ell_{dip,1}$ & $414^{+3}_{-3}$, 17   & $412^{+9}_{-9}$, 4.2 & $414^{+9}_{-9}$, 4,2 & $413^{+7}_{-7}$, 5.5 & $414^{+3}_{-3}$, 18   \\
$\ell_{dip,2}$ &$697^{+41}_{-41}$, 1.0    & $697^{+28}_{-28}$, 1.1  & $698^{+27}_{-27}$, 1.2  & $647^{+20}_{-20}$, 1.6 & $690^{+21}_{-21}$, 1.7   \\
$\ell_{dip,3}$ &  &                         & $1105^{+36}_{-36}$, 2.0    & $1061^{+50}_{-50}$, 1.2  & $1052^{+31}_{-31}$, 1.7  \\
$\ell_{dip,4}$ &  &                         & ($1384^{+17}_{-17}$, 2.4)    & ($1324^{+21}_{-21}$, 2.9)  & ($1324^{+21}_{-21}$, 2.9)  \\
\hline
${\cal C}_{dipk,1}$ & $1619^{+35}_{-35}$ & $1692^{+143}_{-143}$ & $1689^{+122}_{-122}$  & $1742^{+107}_{-107}$ & $1643^{+32}_{-32}$ \\
${\cal C}_{dip,2}$ & $1450^{+234}_{-234}$   & $1889^{+203}_{-203}$  & $1800^{+162}_{-162}$ & $1943^{+132}_{-132}$ & $1768^{+101}_{-101}$  \\
${\cal C}_{dip,3}$ & &                        & $862^{+231}_{-231}$   & $1005^{+218}_{-218}$ & $908^{+187}_{-187}$   \\
${\cal C}_{dip,4}$ & &                        & $115^{+284}_{-284}$   & $468^{+119}_{-119}$   & $448^{+114}_{-114}$   \\
\hline
\end{tabular}
\end{center}
\end{table}

Table~\ref{tab:Lpk} is another approach to peak/dip detection: given a
class of theoretical models with a sequence of peaks and dips, the
statistical distribution of positions and amplitudes can be predicted
by ensemble-averaging over the full probability, the multidimensional
likelihood.

\begin{table}
\caption{Peak/dip locations $\ell_{pk/dip,j}$ and heights ${\cal
C}_{pk/dip,j}$ (in $\mu K^2$) determined by ensemble-averages over
the ${\cal C}_\ell$ database and the weak prior. {\small\rm (Use of the weak prior
allows large movement of peak locations associated with the geometry,
hence is preferable to more restrictive priors for this application. 
These numbers should be contrasted with the `model-independent'
numbers of table~\ref{tab:Lpkslider}. The comoving sound speed, $r_s$,
and damping scale, $R_D$, and their associated angular scales, $\pi
\ell_s$ and $\ell_D$, are also shown.  Database numbers using this
method were given in \deBpkdip01, \Pearson02 and \Ruhl02\ and the
structural parameters for the \junzerotwo\  data were given in \Sievers02.  The results
here compare well with those given by Spergel \et\ (2003) using the
MCMC method: $220 \pm 0.9$ and $535 \pm 2$ for the first two
peaks. The WMAP team also give $r_s = 144 \pm 4 \mpc$ for power-law
$\Lambda$CDM models, and $r_s = 147 \pm 2 \mpc$ using all of the data
and a running index. For $\pi \ell_s$ they get $299 \pm 2$ and $301
\pm 1$, respectively.  When we do WMAP-only, we get similar values to
\marchzerothree\  for the peak locations and amplitudes, but with larger errors
beyond the third for the peaks that WMAP does not cover - errors more
similar to those of the \janzerothree\  data.)}}
\label{tab:Lpk}
\begin{center}
\begin{tabular}{|l|lllll|}
\hline
   & Jan00 & Jan02 & Jun02 & Jan03 &  Mar03  \\
\hline
$\pi \ell_s$ & $292^{+31}_{-28}$ & $299.7^{+3.5}_{-3.5}$ & $300.7^{+3.5}_{-3.5}$ & $299.6^{+2.7}_{-2.7}$ & $300.2^{+1.0}_{-1.0}$ \\
$r_s \, (\mpc)$ & $115^{+24}_{-20}$ & $144^{+7}_{-7}$ & $144^{+6}_{-6}$ & $144.3^{+4.8}_{-4.6}$ & $145.7^{+2.7}_{-2.6}$ \\
$\ell_D$ & $1218^{+304}_{-243}$ & $1345^{+18}_{-18}$ & $1348^{+15}_{-15}$ & $1344^{+14}_{-14}$ & $1353^{+5.4}_{-5.4}$ \\
$R_{D} \, (\mpc )$  & $8.8^{+2.9}_{-2.2}$ & $10.2^{+0.50}_{-0.48}$ & $10.2^{+0.40}_{-0.39}$ & $10.2^{+0.36}_{-0.35}$ & $10.3^{+0.21}_{-0.20}$ \\
\hline
$\ell_{pk,1}$ & $226^{+22}_{-20}$  & $220^{+3}_{-3}$ & $219^{+3}_{-3}$ & $220^{+2}_{-2}$ & $220^{+1}_{-1}$  \\
$\ell_{pk,2}$ & $598^{+153}_{-122}$ & $534^{+13}_{-12}$  & $534^{+12}_{-12}$  & $533^{+13}_{-13}$ & $534^{+2}_{-2}$   \\
$\ell_{pk,3}$ & $881^{+208}_{-168}$ & $810^{+18}_{-18}$  & $812^{+18}_{-17}$ & $810^{+17}_{-17}$  & $812^{+2}_{-2}$  \\
$\ell_{pk,4}$ & $1143^{+289}_{-231}$ & $1124^{+28}_{-28}$ & $1127^{+29}_{-29}$ & $1124^{+25}_{-25}$ & $1125^{+4}_{-4}$  \\
$\ell_{pk,5}$ & $1395^{+444}_{-337}$ & $1420^{+31}_{-31}$ & $1424^{+32}_{-32}$  & $1420^{+28}_{-27}$ & $1423^{+4}_{-4}$ \\
\hline
${\cal C}_{pk,1}$ & $5372^{+690}_{-611}$ & $5442^{+288}_{-274}$ & $5487^{+266}_{-254}$  & $5346^{+177}_{-171}$ & $5551^{+35}_{-34}$ \\
${\cal C}_{pk,2}$ & $2114^{+1841}_{-984}$ & $2495^{+113}_{-108}$ & $2502^{+98}_{-94}$ & $2483^{+80}_{-78}$ & $2486^{+28}_{-28}$  \\
${\cal C}_{pk,3}$ & $1372^{+2678}_{-908}$ & $2482^{+202}_{-187}$ & $2478^{+169}_{-158}$ & $2461^{+143}_{-135}$ & $2454^{+55}_{-54}$   \\
${\cal C}_{pk,4}$ & $836^{+1123}_{-480}$  & $1220^{+104}_{-96}$ & $1203^{+90}_{-84}$  & $1218^{+76}_{-71}$& $1214^{+25}_{-24}$  \\
${\cal C}_{pk,5}$ & $498^{+949}_{-327}$ & $811^{+108}_{-95}$  & $792^{+95}_{-85}$ & $809^{+81}_{-73}$  & $808^{+24}_{-23}$   \\
\hline
$\ell_{dip,1}$ & $393^{+54}_{-48}$  & $412^{+5}_{-5}$ & $413^{+5}_{-5}$ & $412^{+4}_{-4}$ & $409^{+1}_{-1}$  \\
$\ell_{dip,2}$ & $507^{+162}_{-123}$ & $671^{+28}_{-27}$  & $673^{+23}_{-22}$  & $670^{+10}_{-10}$ & $672^{+2}_{-2}$   \\
$\ell_{dip,3}$ & $830^{+242}_{-188}$ & $1012^{+31}_{-30}$  & $1014^{+28}_{-27}$ & $1009^{+15}_{-15}$  & $1010^{+4}_{-4}$  \\
$\ell_{dip,4}$ & $1097^{+313}_{-244}$ & $1306^{+36}_{-35}$ & $1308^{+33}_{-32}$ & $1302^{+18}_{-18}$ & $1305^{+4}_{-4}$  \\
\hline
${\cal C}_{dipk,1}$ & $1932^{+1020}_{-668}$ & $1638^{+78}_{-74}$ & $1637^{+70}_{-70}$  & $1627^{+57}_{-55}$ & $1654^{+15}_{-15}$ \\
${\cal C}_{dip,2}$ & $1130^{+1453}_{-634}$ & $1712^{+144}_{-133}$ & $1696^{+119}_{-112}$ & $1707^{+68}_{-66}$ & $1689^{+35}_{-35}$  \\
${\cal C}_{dip,3}$ & $709^{+1469}_{-479}$ & $974^{+80}_{-75}$ & $962^{+68}_{-64}$ & $975^{+44}_{-43}$ & $964^{+19}_{-18}$   \\
${\cal C}_{dip,4}$ & $557^{+834}_{-334}$  & $668^{+82}_{-73}$ & $656^{+70}_{-63}$  & $670^{+43}_{-41}$& $662^{+18}_{-17}$  \\
\hline
\end{tabular}
\end{center}
\end{table}

\subsection{Characteristic scales:  sound crossing, peaks and dips, and damping} \label{sec:LsLDLpk}

A strong first peak followed by a sequence of smaller peaks diminished
by damping in the ${\cal C}_\ell$ spectrum was a long-standing
prediction of adiabatic models (pillar 2). The critical scale
determining the spatial positions of the acoustic peaks in the spectra
of figure~\ref{fig:CLoptmar03} is the (comoving) sound crossing distance
at recombination, $r_s$. The corresponding multipole scale is $\ell_s
\equiv {\cal R}_{dec}/r_s$, where ${\cal R}_{dec}$ is the
angular-diameter distance that maps angles observed at our location to
comoving spatial scales at recombination.

In terms of the comoving distance $\chi_{dec} $ to photon decoupling
(recombination, at redshift $z_{dec}= a_{dec}^{-1}-1$), and the
curvature scale $d_k$, ${\cal R}_{dec}$ is given by
\begin{eqnarray} 
&& {\cal R}_{dec} =\{d_k {\rm sinh} (\chi_{dec}/d_k), \chi_{dec}, d_k
{\rm sin} (\chi_{dec}/d_k)\}, \ 
\end{eqnarray}
where $d_k =3000 |\omega_k|^{-1/2} \mpc$ and 
\begin{eqnarray} 
&& \chi_{dec} = 6000
\mpc \int_{\sqrt{a_{dec}}}^{1} (\omega_m + \omega_Q a^{-3w_Q}
+\omega_k a)^{-1/2}\ d\sqrt{a} \, .  \nonumber 
\end{eqnarray}
The three cases are for negative, zero and positive mean curvature. Thus
the mapping depends upon $\omega_{k}$, $\omega_Q$ and $w_Q$ as well as
on $\omega_m$. 

The sound crossing distance at recombination is
\begin{equation} 
r_s =
{6000\over \sqrt{3}} \mpc \int_{0}^{\sqrt{a_{dec}}} (\omega_m + \omega_{er}
a^{-1})^{-1/2} (1+ \omega_b a/(4\omega_\gamma /3))^{-1/2}\
d\sqrt{a}, 
\end{equation}
where $\omega_\gamma = 2.46 \times 10^{-5}$ is the
photon density and $\omega_{er} = 1.68 \omega_\gamma$ for three species
of massless neutrinos. 

The estimates of $r_s$ and $\pi \ell_s$ have been quite stable over
time (table~\ref{tab:Lpk}). Values are determined by averaging over
the ${\cal C}_\ell$ model space probabilities. Since the $r_s$ are
comoving, the physical sound horizon at decoupling is \ca\ $140
\kpc$.

The angular-diameter-distance relation maps spatial structure at
photon decoupling perpendicular to the line-of-sight with transverse
wavenumber $k_\perp$ to angular structure, through $\ell = {\cal
R}_{dec} k_\perp$. Converting peaks in $k$-space into peaks in
$\ell$-space is complicated by three-dimensional to 2D projection
effects over the finite width of decoupling and also by the influence
of sources other than sound oscillations such as Doppler terms.

The peak locations $\ell_{pk,j}$ in table~\ref{tab:Lpk} are obtained
by forming $\exp<\ln \ell_{pk,j}>$, where the average and variance of
$\ln \ell_{pk,j}$ are determined by integrating over the
probability-weighted ${ \cal C}_\ell$ database.  These peak locations
accord reasonably well with $\ell_{pk,j} \approx j f_j \pi \ell_s$,
where the numerically estimated constant $f_j$ $\approx 0.75$ for
the first peak, approaching unity for higher ones. (There are small
$n_s-1$ corrections to $\ln f_j$.)  The interleaving dips are also
shown.  Dip locations are well determined by replacing $j$ by $j+1/2$,
with slightly different $f_j$ factors.

For fixed $\omega_b$ and $\omega_m$, constant $\ell_s$ lines in the
$\Omega_k$--$\Omega_\Lambda$ and $w_Q$--$\Omega_Q$ planes look rather
similar to contour lines determined from the data (see figures~5 and 7 in
\npbmrst02). This degeneracy (\npdegeneracies) among these
parameters would be exact except for the integrated Sachs-Wolfe
effect. However, lines of constant $H_0$ are different in that space
and break the degeneracy, only weakly for the weak
prior but more so when the HST-h or the SN1 prior on deceleration is
imposed. The degeneracy in the $\Omega_k$--$\Omega_\Lambda$ plane is also
broken when LSS information is added, though less so in the
$w_Q$--$\Omega_Q$ plane.

Also evident in the spectra in figure~\ref{fig:CLoptmar03} and the
diminishment of the peak heights in table~\ref{tab:Lpk} is the damping
tail, an overall decline due to the shear viscosity and the finite
width of the region over which hydrogen recombination occurs. Scales
for both can be estimated analytically and are similar. Values for the
comoving scale $R_D$ and an associated angular damping scale
$\ell_D\equiv {\cal R}_{dec}/R_D$ are given in
table~\ref{tab:Lpk}. The physical scale $R_D/(1+z_{dec})$ at
recombination is therefore about 10 kpc. (The decline in ${\cal
C}_\ell$ is parameterized by a $\exp[-(\ell/\ell_D)^{m_D}]$ multiplier
(Hu and White 1996). We find $m_D$ = $1.27^{+0.008}_{-0.008}$ for
\janzerothree\  and $1.27^{+0.003}_{-0.003}$ for \marchzerothree.)

 The $\omega_b$ dependence in $r_s$ would lead to a degeneracy with
other parameters in terms of peak/dip positions. However, relative
peak/dip heights are extremely significant for parameter estimation as
well, and this breaks the degeneracy. For example, increasing
$\omega_b$ beyond the nucleosynthesis (and CMB) estimate leads to a
diminished height for the second peak that is not in accord with the
data. \wmapP\ used the peak/dip parameters to estimate directly the 
cosmic parameters.
 
\section{Cosmic parameter estimations} \label{sec:params}

\subsection{Evolution of marginalized cosmic parameters} \label{sec:margparam}

\begin{figure}
\hspace*{-8mm}
\includegraphics[width=5.8in]{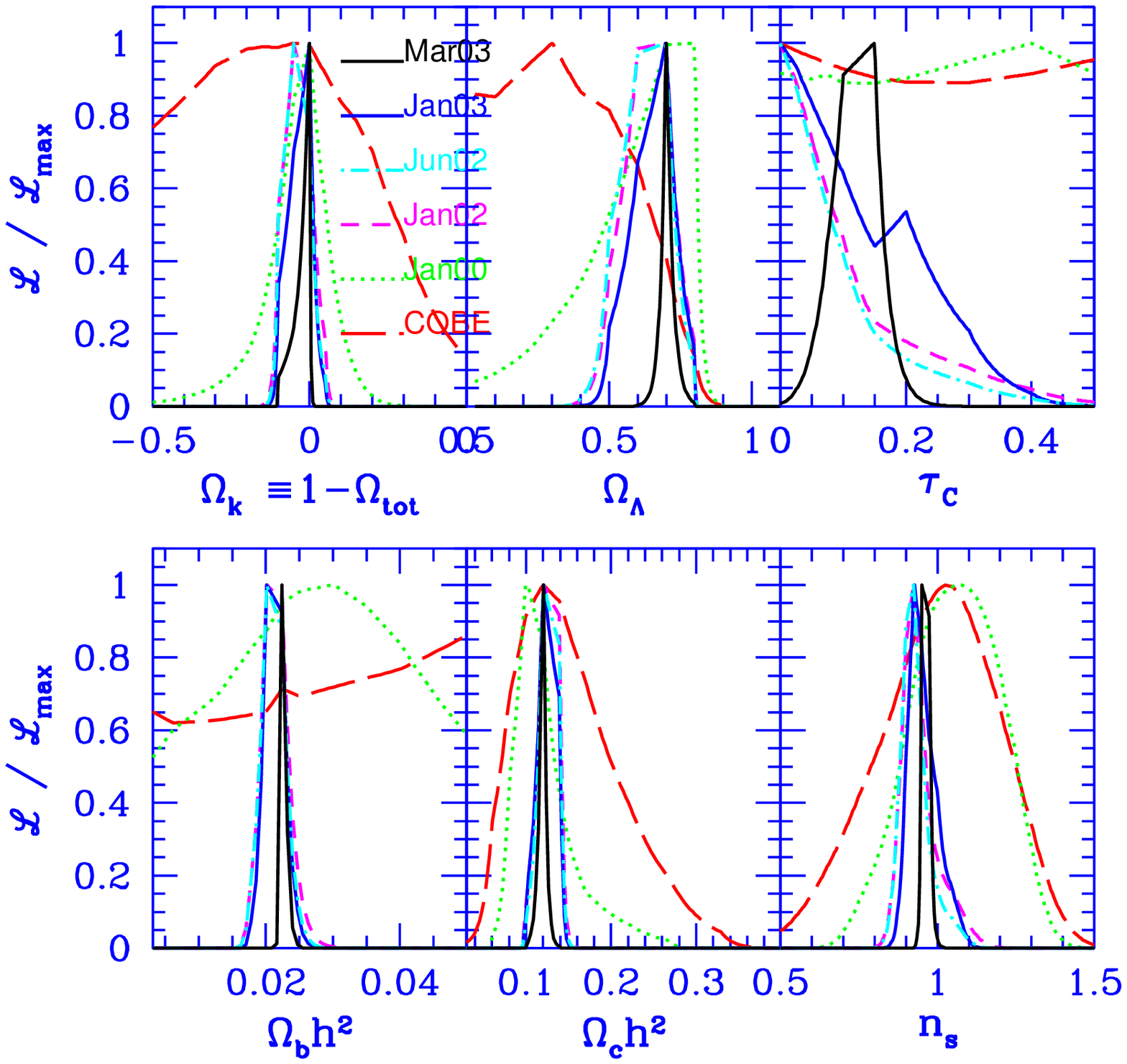}
\caption{1D likelihood curves for the weak+flat+LSS prior show that
the parameters were in good agreement but also show how much WMAP has
sharpened the picture.}
\label{fig:1DwkflatLSS}
\end{figure}

Table~\ref{tab:exptparams} and figure~\ref{fig:1DwkflatLSS} show the
evolution of cosmic parameter estimations from projected 1D likelihood
curves, using the ${\cal C}_\ell$ database used in \Sievers02,
\Ruhl02 and \Goldstein02, an extension of that used in \lange00,
\jaffe00 and \nett01\ to reflect the growing precision. The ranges quoted
are Bayesian 50\% values and the errors are 1$\sigma$, obtained after
projecting (marginalizing) over all other parameters. For derived
variables such as $h$ amd age, ensemble-averaged means and variances
are used. The WMAP addition punches out the detection in a remarkable
way.

The MCMC parameter estimates for weak and weak+flat priors in
table~\ref{tab:exptparamsMCMC} agree well with the entries in
table~\ref{tab:exptparams}: about as good as one might expect given
the differences. In particular, WMAP with MCMC includes the
$\tau_C$ constraint by explicit calculation of TE likelihoods, while
the database either lets the TT data alone decide or uses the
broader-than-Gaussian $\tau_C$ prior to reflect the detection. The
\marchzerothree\  results in the table include this prior, but most parameters
are insensitive to it: even $\sigma_8$ does not migrate that much.

The Spergel \et\ (2003) MCMC four-chain results for flat power law
$\Lambda$CDM models for WMAP-only are quite similar to what we
obtain. Their WMAP-ext consists of WMAP plus cut verions of the \Acbar\ 
and CBI one-year power spectra. Our \marchzerothree\  set contains more data and
allows more overlap.  Their analogue of LSS is to use the 2dFRS power
spectrum.  This assumes a linear scale-independent bias. It is a more
stringent prior on shape than our $\Gamma$ constraint.  They also add
(small-scale) Ly$\alpha$ forest information to extend the $k$-space
coverage.  The forest data could be sensitive to gastrophysical
complications, and we would hesitate to construct any but a very weak
prior for it at this time.

\begin{figure}
\includegraphics[width=5.3in]{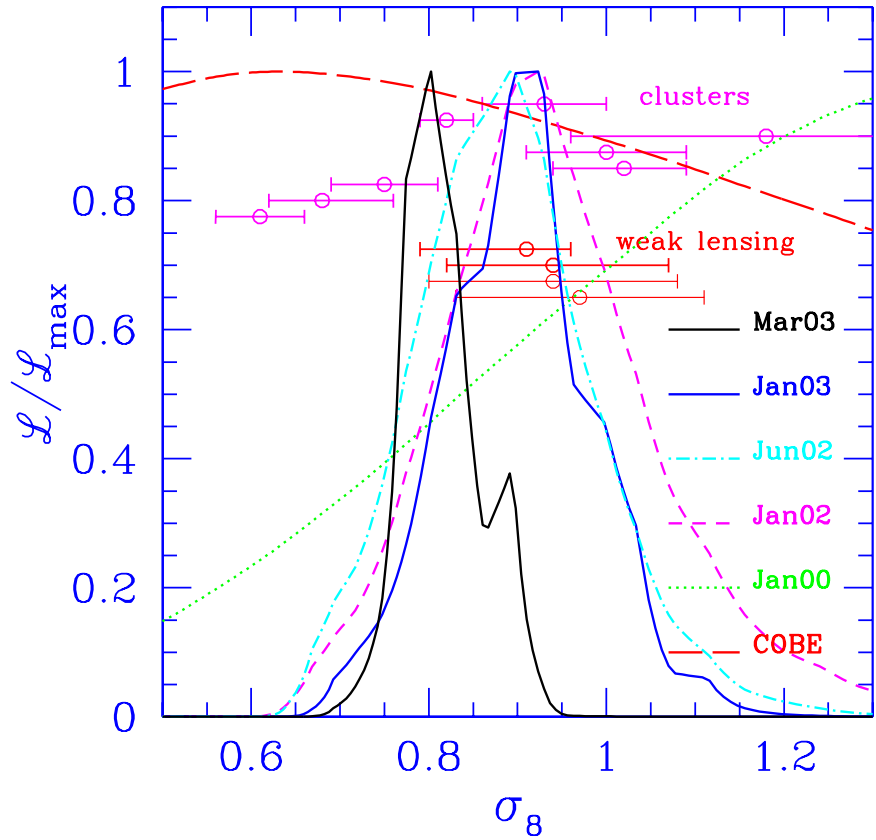}
\caption{Evolution of $\sigma_8$ determinations for the weak+flat
prior, with the $\tau_C$ prior applied as well for the \marchzerothree\  data.
These are contrasted with the estimates from weak lensing and clusters
which the LSS prior encompasses (\npSievers02; \npBond02).  The MCMC
results give distributions that extend slightly more to higher
$\sigma_8$. }
\label{fig:sig8evoln}
\end{figure}

We also find good agreement with Contaldi, Hoekstra and Lewis (2003),
who applied Cosmomc to the WMAP+\Acbar+CBI data and the Red Cluster
Survey (RCS) weak-lensing data. Our LSS prior was designed to
encompass $\sigma_8$ constraints from weak lensing displayed in
figure~\ref{fig:sig8evoln}, and RCS is one of those entries. 

Table~\ref{tab:exptparams} shows that with just the weak prior, there
are strong detections for $\Omega_{tot}$, $\omega_b$,
$\omega_{cdm}$ and $n_s$. The HST-h prior helps to determine
$\Omega_\Lambda$ better than the weak prior, because it breaks more
strongly the $\Omega_{k}$--$\Omega_\Lambda$ near-degeneracy. SN1
breaks it even more strongly, and the HST-h, weak+SN1 and weak+LSS
results are all compatible, showing there is no very strong tension
among these priors.

\begin{table}
\caption{Cosmological parameter values from the ${\cal C}_\ell$ database and their 1$\sigma$ errors are
shown, determined after marginalizing over the other six cosmological
and the various experimental parameters. {\small\rm (In all cases, the weak prior
($0.45 \le {\rm h} \le 0.9$, age older than $10$ Gyr) is applied, and $w_Q$ is
fixed at $-1$, the cosmological constant case.  The first set allow
$\Omega_k$ to vary, the rest fix it at zero. The LSS prior agrees with
weak lensing and redshift survey results, and agrees with most of the
cluster determinations. The parameters are very stable if extra
`prior' probabilities for LSS are included, or if the HST range for
$h$ is used, or if SN1 data are included. Allowing $w_Q$ to vary
yields quite similar results. Although the optimal spectra include the
\janzerozero\  data in the subsequent mixes in figure~\ref{fig:CLoptevoln}, these
parameter estimates do not. (Their inclusion has little effect.))}}
\label{tab:exptparams}
\begin{center}
\begin{tabular}{|l|lllll|}
\hline
database & Jan00 & Jan02 & Jun02 & Jan03 &  Mar03 \\
\hline
weak &  &  &  & &   \\
\hline $\Omega_{tot}$           & $1.06^{+.16}_{-.10}$ &
$1.035^{+.043}_{-.046}$ & $1.038^{+.040}_{-.042}$ & $1.034^{+.039}_{-.036}$ & $1.015^{+.063}_{-.015}$  \\
$\Omega_b{\rm h}^2$             & $.0339^{+.0443}_{-.0246}$ &
$.0222^{+.0025}_{-.0021}$ & $.0221^{+.0024}_{-.0020}$ & $.0221^{+.0023}_{-.0018}$ & $.0233^{+.0013}_{-.0013}$  \\
$\Omega_{cdm}{\rm h}^2$ & $.198^{+.088}_{-.080}$ &
$.130^{+.031}_{-.028}$ & $.124^{+.026}_{-.025}$ & $.125^{+.021}_{-.022}$  & $.111^{+.010}_{-.010}$  \\
$n_s$            & $1.218^{+.135}_{-.163}$ & $0.949^{+.083}_{-.049}$ 
& $0.938^{+.077}_{-.042}$ &
$0.961^{+.081}_{-.047}$ & $0.978^{+.025}_{-.020}$ \\
$\Omega_{\Lambda}$ & $0.34^{+.28}_{-.24}$ & $0.52^{+.17}_{-.20}$ & $0.53^{+.17}_{-.19}$ &
$0.57^{+.14}_{-.19}$ & $0.73^{+.06}_{-.10}$  \\
$h$ & $0.60 \pm  0.11$ & $0.56 \pm 0.10$  & $0.55 \pm 0.10 $ &
$0.57 \pm 0.10 $ & $0.68 \pm 0.13 $ \\
$age$ & $12.9 \pm 2.1$ &  $15.0 \pm 1.4$ &  $15.1 \pm 1.3$ & $14.9 \pm 1.2$
 & $14.2 \pm 1.3$  \\
$\Omega_m$ & $0.72 \pm 0.29$ & $0.53 \pm 0.19$ &$0.51 \pm 0.18$ &
$0.49 \pm 0.18$ & $0.33 \pm 0.19$  \\
\hline
flat+weak  &   &  &  &  &   \\
\hline 
$\Omega_b{\rm h}^2$ & $.0309^{+.0259}_{-.0192}$ &
$.0213^{+.0020}_{-.0017}$ & $.0213^{+.0019}_{-.0017}$ & $.0216^{+.0016}_{-.0015}$ 
 & $.0234^{+.0013}_{-.0013}$ \\
$\Omega_{cdm}{\rm h}^2$& $.176^{+.083}_{-.068}$ & $.141^{+.025}_{-.025}$ & $.135^{+.024}_{-.022}$
& $.131^{+.020}_{-.019}$ & $.111^{+.010}_{-.010}$\\
$n_s$ & $1.141^{+.131}_{-.153}$ & $0.931^{+.060}_{-.040}$ & $0.924^{+.050}_{-.036}$ &
$0.951^{+.062}_{-.039}$ &  $0.979^{+.019}_{-.018}$  \\
$\Omega_{\Lambda}$& $0.47^{+.23}_{-.30}$ & $0.54^{+.18}_{-.26}$ & $0.58^{+.16}_{-.26}$ &
$0.64^{+.11}_{-.16}$ & $0.75^{+.04}_{-.05}$ \\
$h$ & $0.66 \pm 0.12$ & $0.61 \pm 0.10$ & $0.63 \pm 0.10$ &
$0.66 \pm 0.09 $ & $0.75 \pm 0.05 $  \\
$age$ & $12.7 \pm 1.9$ & $13.9 \pm 0.5$ & $14.0 \pm 0.5$ &
$13.8 \pm 0.34$ &  $13.5 \pm 0.14$  \\
$\Omega_m$ & $0.55 \pm 0.24$ & $0.49 \pm 0.20$ & $0.45 \pm 0.20$ &
$0.38 \pm 0.15 $ & $0.24 \pm 0.05$  \\
$\sigma_{8}$ & $1.23^{+.47}_{-.47}$ & $0.92^{+.12}_{-.12}$ & $0.88^{+.10}_{-.11}$ &
$0.90^{+.09}_{-.09}$ & $0.81^{+.06}_{-.04}$ \\
$\Omega_m h$ & $0.34 \pm 0.13$ & $0.28 \pm 0.08 $ & $0.26 \pm 0.08 $ &
$0.24 \pm 0.07 $ & $0.18 \pm 0.03$  \\
$\Gamma$ & $0.28 \pm 0.12$ & $0.24 \pm 0.08$ &
$0.23 \pm 0.07 $ & $0.20 \pm 0.06$  & $0.14 \pm 0.03$  \\
\hline
LSS+flat+weak  &  & &  &  & \\
\hline 
$\Omega_b{\rm h}^2$& $.036^{+.03}_{-.02}$ &
$.0217^{+.0019}_{-.0018}$ & $.0215^{+.0018}_{-.0017}$ & $.0217^{+.0015}_{-.0015}$ &
$.0228^{+.0013}_{-.0013}$ \\
$\Omega_{cdm}{\rm h}^2$& $.11^{+.04}_{-.03}$ & $.128^{+.011}_{-.012}$ & $.128^{+.011}_{-.011}$
& $.126^{+.012}_{-.012}$ & $.121^{+.010}_{-.010}$\\
$n_s$ & $1.09^{+.15}_{-.16}$ & $0.93^{+.07}_{-.04}$ & $0.93^{+.05}_{-.04}$ &
$0.950^{+.067}_{-.037}$  & $0.965^{+.013}_{-.013}$  \\
$\Omega_{\Lambda}$& $0.61^{+.09}_{-.38}$ & $0.64^{+.08}_{-.09}$ & $0.63^{+.08}_{-.10}$ &
$0.66^{+.07}_{-.09}$ & $0.70^{+.05}_{-.05}$ \\
$h$ & $0.67 \pm 0.13$ & $0.65 \pm 0.07 $ & $0.65 \pm 0.06 $ &
$0.66 \pm 0.06 $ & $0.69 \pm 0.01 $  \\
$age$ & $13.7 \pm 1.9$ & $13.8 \pm 0.5$ & $13.9 \pm 0.5$  & 
$13.8 \pm 0.28 $ & $13.7 \pm 0.03$   \\
$\Omega_m$ & $0.40 \pm 0.19$ & $0.36 \pm 0.09$ &$0.37 \pm 0.09$ &
$0.35 \pm 0.08 $  & $0.30 \pm 0.011$  \\
$\sigma_{8}$& $0.85^{+.26}_{-.22}$ & $0.88^{+.09}_{-.08}$ & $0.86^{+.08}_{-.07}$ &
$0.89^{+.06}_{-.07}$ & $0.86^{+.04}_{-.04}$ \\
$\Omega_m h$ & $0.25 \pm 0.10$ & $0.23 \pm 0.04 $ &$0.24 \pm 0.04 $ &
$0.23 \pm 0.04 $ & $0.206 \pm 0.006$  \\
$\Gamma$ & $0.18 \pm 0.05$ & $0.20 \pm 0.04$ &$0.20 \pm 0.04$ &
$0.19 \pm 0.033 $ & $0.174 \pm 0.006$  \\
\hline
\end{tabular}
\end{center}
\end{table}

\begin{table}
\caption{MCMC determinations of cosmic parameters should be compared
with those of the ${\cal C}_\ell$ database given in
table~\ref{tab:exptparams}. {\small\rm (Means and standard deviations
are given here, rather than the Bayesian 50\%, with 16\% and 84\%
`1$\sigma$' errors. In addition, for the \marchzerothree\  data, the Bayesian
results using the 899 WMAP points in the database with the
$\tau_C$-prior are shown in the last column. Note how $\sigma_8$
increases when the logarithmic running of the spectral slope is
allowed, a manifestation of the high correlation of $\sigma_8$ and
$\tau_C$ with $dn_s/d\ln k$ as well as $n_s$. The values are therefore
sensitive to the priors imposed.)} }
\label{tab:exptparamsMCMC}
\begin{center}
\begin{tabular}{|l|lllll|}
\hline
MCMC    & Jan02 & Jun02 & Jan03 &  Mar03 & Mar03(899db) \\
\hline
weak &  &  &  & &   \\
\hline 
$\Omega_{tot}$             &
$1.024^{+.041}_{-.041}$ & $1.030^{+.039}_{-.039}$ & $1.039^{+.037}_{-.037}$ & $1.050^{+.032}_{-.032}$  & $1.016^{+.08}_{-.03}$  \\
$\Omega_b{\rm h}^2$              &
$.0215^{+.0017}_{-.0017}$ & $.0212^{+.0015}_{-.0015}$ & $.0214^{+.0013}_{-.0013}$ & $.0222^{+.0007}_{-.0007}$  & $.0227^{+.0013}_{-.0013}$\\
$\Omega_{cdm}{\rm h}^2$ &  
$.145^{+.026}_{-.026}$ & $.135^{+.023}_{-.023}$ & $.131^{+.017}_{-.017}$  & $.120^{+.009}_{-.009}$ & $.112^{+.010}_{-.010}$  \\
$n_s$              & $0.995^{+.063}_{-.063}$ 
& $0.976^{+.053}_{-.053}$ &
$0.965^{+.039}_{-.039}$ & $0.950^{+.017}_{-.017}$ & $0.960^{+.016}_{-.013}$\\
$\Omega_{\Lambda}$   & $0.45^{+.18}_{-.18}$ & $0.48^0{+.16}_{-.16}$ &
$0.51^{+.15}_{-.15}$ & $0.55^{+.11}_{-.11}$ & $0.71^{+.06}_{-.30}$ \\
$h$    & $0.56 \pm 0.11$  & $0.55 \pm 0.10$ &
$0.56 \pm 0.11 $ & $0.55 \pm 0.09$ &  $0.61 \pm 0.14 $ \\
$age$ &    $14.4 \pm 1.3$ &  $14.9 \pm 1.3$ & $14.9 \pm 1.2$
 & $15.3 \pm 1.0$  & $14.8 \pm 1.5$  \\
$\Omega_m$  & $0.58 \pm 0.19$ & $0.55 \pm 0.17$ &
$0.54 \pm 0.18$ & $0.51 \pm 0.14$ & $0.45 \pm 0.22$ \\
\hline
flat+weak  &   &  &  &  &   \\
\hline 
$\Omega_b{\rm h}^2$ &  
$.0226^{+.0026}_{-.0026}$ & $.0219^{+.0024}_{-.0024}$ & $.0219^{+.0018}_{-.0018}$ 
 & $.0230^{+.0011}_{-.0011}$ & $.0228^{+.0013}_{-.0013}$ \\
$\Omega_{cdm}{\rm h}^2$  & $.140^{+.026}_{-.026}$ & $.132^{+.024}_{-.024}$
& $.128^{+.018}_{-.018}$ & $.117^{+.010}_{-.010}$ & $.116^{+.010}_{-.010}$ \\
$n_s$   & $1.02^{+.087}_{-.087}$ & $0.994^{+.073}_{-.073}$ &
$0.973^{+.045}_{-.045}$ &   $0.967^{+.029}_{-.029}$ & $0.965^{+.015}_{-.013}$ \\
$\Omega_{\Lambda}$ & $0.56^{+.20}_{-.20}$ & $0.58^{+.19}_{-.19}$ &
$0.65^{+.13}_{-.13}$ & $0.72^{+.05}_{-.05}$ & $0.73^{+.05}_{-.05}$ \\
$h$   & $0.64 \pm 0.11$ & $0.64 \pm 0.10$ &
$0.67 \pm 0.08 $ & $0.71 \pm 0.05 $  & $0.72 \pm 0.05 $ \\
$age$   & $13.6 \pm 0.6$ & $13.9 \pm 0.5$ &
$13.6 \pm 0.40$ &  $13.6 \pm 0.22$ & $13.6 \pm 0.12$\\
$\Omega_m$  & $0.44 \pm 0.20$ & $0.42 \pm 0.19$ & 
$0.35 \pm 0.13 $ & $0.28 \pm 0.05$ &$0.27 \pm 0.05$   \\
$\sigma_{8}$   & $1.05^{+.20}_{-.20}$ & $0.94^{+.13}_{-.13}$ &
$0.89^{+.09}_{-.09}$ & $0.85^{+.06}_{-.06}$ & $0.83^{+.05}_{-.06}$ \\
\hline
\hline
flat+weak  & $dn_s/d\ln k$  &  &  &  &   \\
\hline 
$\Omega_b{\rm h}^2$ &  
$.0211^{+.0027}_{-.0027}$ & $.0207^{+.0024}_{-.0024}$ & $.0207^{+.0020}_{-.0020}$ 
 & $.0229^{+.0018}_{-.0018}$ & \\
$\Omega_{cdm}{\rm h}^2$  & $.154^{+.028}_{-.028}$ & $.147^{+.026}_{-.026}$
& $.149^{+.025}_{-.025}$ & $.121^{+.016}_{-.016}$ &  \\
$n_s(k_n)$   & $0.920^{+.119}_{-.119}$ & $0.897^{+.101}_{-.101}$ &
$0.874^{+.075}_{-.075}$ &   $0.924^{+.059}_{-.059}$ &  \\
$-dn_s(k_n)/d\ln k$   & $0.098^{+.060}_{-.060}$ & $0.101^{+.056}_{-.056}$ &
$0.091^{+.045}_{-.045}$ &   $0.083^{+.033}_{-.033}$ &  \\
$\Omega_{\Lambda}$ & $0.44^{+.23}_{-.23}$ & $0.45^{+.23}_{-.23}$ &
$0.48^{+.21}_{-.21}$ & $0.70^{+.10}_{-.10}$ &  \\
$h$   & $0.59 \pm 0.10$ & $0.58 \pm 0.10$ &
$0.60 \pm 0.10$ & $0.71 \pm 0.08 $  &  \\
$age$   & $13.8 \pm 0.6$ & $14.0 \pm 0.5$ &
$13.8 \pm 0.42$ &  $13.5 \pm 0.36$ & \\
$\Omega_m$  & $0.56 \pm 0.23$ & $0.55 \pm 0.23$ & 
$0.52 \pm 0.21 $ & $0.30 \pm 0.10$ &   \\
$\sigma_{8}$   & $1.20^{+.24}_{-.24}$ & $1.13^{+.22}_{-.22}$ &
$0.99^{+.13}_{-.13}$ & $0.96^{+.08}_{-.08}$ &  \\
\hline
\end{tabular}
\end{center}
\end{table}

The precision of WMAP invites exploration of larger parameter spaces.
A first issue is whether $n_s$ varies.  When \wmapS\ added the
high-$\ell$ \Acbar\  and CBI data to make their WMAP-ext dataset, they
found lower $n_s$, and further adding the 2dFRS data and the
small-scale data from the Lyman alpha forest exacerbated the issue.
We find the same result with the database with MCMC: $n_s$ changes by
about 0.03 from WMAP-only to the \marchzerothree\  set for the weak+flat prior,
and by a further 0.015 with LSS added.

Modelling $n_s$ variation with a logarithmic correction, \wmapS\ get
$n_s$ varying from 1.2 to 0.93. (See also Bridle \et\ 2003, who show
that dropping low-$\ell$ modes reduces the significance of running
index detection.) The lower panel of table~\ref{tab:exptparamsMCMC}
shows how the parameters change relative to the fixed $n_s$ model when
this extra parameter is introduced, for the weak+flat prior and for
the various datasets. We restricted $dn_s /d\ln k$ to lie within the
range $-0.2$-$0.2$. The normalization point for $n_s$ is $k_n^{-1}=20
\hmpc$, and the horizon scale is $\chi_{dec} \approx 0.88
(6000/\Omega_m^{1/2}) \hmpc$ for a $\Lambda$CDM model, so $\ln k_n
\chi_{dec} \sim 6$, hence substantial slope changes are possible for
the $dn_s/d\ln k$ values obtained. Although
table~\ref{tab:exptparamsMCMC} shows there is a preference for a
logarithmically varying slope from January 2002, due to downturns in
the CMB data at low and high $\ell$, it is $\lta 2\sigma$ in
likelihood fall-off even in the March 2003 data. The likelihood maxima
are also smaller than the means given in the table. We also caution
that $dn_s/d\ln k (k_n)$ is correlated with the
$\tau_C$-$\sigma_8$-$n_s$ combination, which results in upward
extension of the 1D distributions for $\tau_C$ and $\sigma_8$. Prior
choices on $\tau_C$ can then move the limits around: the weak+flat
results in table~\ref{tab:exptparamsMCMC} place no restriction beyond
the weak 0.7 upper boundary on $\tau_C$ to conform MCMC to our
database limit.

The WMAP team estimate the gravitational wave
(tensor) contribution to be $< 0.72$ of the scalar component in
amplitude. 

We have considered variations in the dark energy equation of
state. For the database calculations, we used a flat prior and allowed
$w_Q$ to vary from $-1$ to $-0.01$, as in \capp00. We applied it to the
\junzerotwo\  data (and obtained $w_Q < -0.7$ at the 95\% CL for the prior
combination weak+flat+LSS+SN1) (\npbmrst02) and to what was almost the
same as the \janzerothree\  data (for which $w_Q < -0.8$ was obtained)
(Pogosyan \et\ 2003). MCMC calculations yield about the same limits:
less than $-0.70$ at $2\sigma$ for the HST-h+flat+LSS and SN1
prior. We get only slightly better limits for the \marchzerothree\  data (less
than $-0.71$). This constraint is SN1-driven (for weak+SN1 $ w_Q <
-0.68$) rather than HST-h-driven ($w_Q < -0.48$). Although bands of
allowed $h$ do break the angular-diameter-distance degeneracy between
$\Omega_Q$ and $w_Q$, allowed bands for the deceleration parameter
$q_0$ break it more effectively.

Unlike the WMAP team, our weak+flat+LSS prior does not give a good
constraint. Since the shape parameter $\Gamma \sim \omega_m /h$, and
$\omega_m$ is now accurately determined by the CMB data, our shape
constraint approaches a pure $h$ prior. For $\omega_m \sim 0.14$, it
is similar to the weak limits we are already imposing on $h$. The 2dFRS 
data give $\Gamma =0.21 \pm 0.03$ and the SDSS data give $0.19 \pm
0.04$, so the combination would nomimally have an 11\% relative
Gaussian error, the same as for the HST-h prior, suggesting the
constraint would be similar if this strong-LSS were to be imposed.

Why the limits are typically $< - 0.7$ or $ -0.8 $ with SN1 is easy to
understand: the $2\sigma$ SN1 error contour in the $\Omega_Q$--$w_Q$
plane roughly follows a $q_0$ $\approx -1/4$ line, where $q_0 =
(1+3w_Q \Omega_Q)/2$ is the current deceleration parameter. With
$\Omega_Q \gta 0.65$, this gives the -0.7 to -0.8 range for the
limit. The way the CMB comes into this is to restrict the allowed
values of $\Omega_Q$, and it used to be that the data was such that
the LSS prior was needed to do this the first time we went through
this exercise. The addition of WMAP refines $\Omega_Q$ but this only
fine tunes the limit by a small amount.

The other component to our LSS prior, the amplitude constraint
$\sigma_8$, has had a significant impact in constraining models.  The
\junzerotwo\  data for the weak + flat prior give $\sigma_8= 0.88 \pm 0.11$
and when LSS is added, $0.86 \pm 0.08$ (\npBond02). For the \marchzerothree\  data, the MCMC calculation
with the weak+flat prior gives $0.85 \pm 0.06$, the database calculation 
with the weak+flat+$\tau_C$ prior gives $0.83 \pm 0.06$ using the
899-point WMAP data and $0.81 \pm 0.06$ using the compressed 98-band
WMAP data. Adding the LSS prior gives $0.86 \pm 0.05$, $0.85 \pm 0.04$ and $0.86 \pm 0.04$,
respectively. (These $\sigma_8$ differences, plus $\sim 0.01$
excursions in $n_s$, represent the biggest deviations between
compressed and uncompressed WMAP estimations; and the up and down
1$\sigma$ and 2$\sigma$ limits are close.)

The SZ effect breaks the $\sigma_8$-$\tau_C$ degeneracy. The SZ power
spectrum is found to scale as $\sigma_8^7$ about a $\Lambda$CDM model
with $\sigma_8 =0.9$ (\eg \npBond02). This high non-linearity means that
it could be an excellent way to estimate $\sigma_8$. For an SZ
explanation to work for the (recalibrated) CBI+BIMA+\Acbar\ data seems to require
$\sigma_8 \approx 0.94^{+0.08}_{-0.16}$ with 1$\sigma$ errors which
include the non-Gaussian nature of the SZ effect in the small patches
that the CBI deep and BIMA measurements probe (\npGoldstein02). These SZ
distributions overlap with those found for the CMB and CMB+LSS data,
but the jury is still out on whether this is the explanation.

The calculation of the SZ angular power spectrum requires detailed
hydrodynamical simulations that properly take into account cluster
structure, pressure profiles, heating/cooling, \etc\ There are still
uncertainties as to the role of feedback, and much work is needed on
the theoretical side as well as the observational side to turn this
technique into a high-precision tool.

\subsection{Parameter eigenmodes and degeneracy breaking} \label{sec:eigen}

To help break parameter near-degeneracies via CMB alone, one
can decrease $\Delta \ell$ of the bands with fixed error, as 
\Archeops\ and WMAP did.  We can also extend the $\ell$ range,
which CBI, \Acbar\ and the VSA have done.  Planck will do both. 

Polarization helps, by breaking the $\tau_C$-amplitude near-degeneracy
and also by breaking the $n_s(k)$ degeneracy that arises if we let the
index have full functional freedom. The EE power near the peak will be
well probed by many of the polarization experiments mentioned in
\S~\ref{sec:cmbdata}\ref{sec:pol}.  

Degeneracy breaking also arises when non-CMB data are added, as the HST-h,
SN1a, and LSS examples show. Higher order statistics can also help,
\eg skewness/kurtosis. 

Forecasting how proposed experiments would improve on cosmic parameter
errors became, and remains, quite a cottage industry. As mentioned in
\S~\ref{sec:GUSparam}\ref{sec:pareigen}, a figure of merit is how
many parameter eigenmodes will be determined to a specific precision
level, \ie the number of modes with $\sigma_\alpha$ below some number,
taken here to be 0.l and 0.01, respectively.  

How well did we do?  The forecast for Boomerang+DMR for 1.8\% of the
sky with four 150 GHz bolometers gave four out of nine linear combinations
should be determined to $\pm 0.1$ accuracy. This is what was obtained
in the full analysis of \nett01. (For the forecasts, $\omega_{hdm}$
and $n_t$ were included in the parameter mix, $w_Q$ was frozen at
$-1$, making nine.) The forecast for WMAP with two years of data was 6/9 to
$\pm 0.1$ and 3/9 to $\pm 0.01$. For one year of WMAP-only data and the
weak prior, 2/7 are determined to $\pm 0.01$, 5/7 to $\pm 0.1$;
adding the $\tau_C$ prior increase this to 6/7. (For Planck, 5/9 to
$\pm 0.01$ accuracy are predicted.)

The \marchzerothree\  eigenmodes for the weak prior are: to $\pm 0.004$, a
primarily $\Omega_k$-$\Omega_\Lambda$ combination, related to $\pi
\ell_s$; to $\pm 0.008$, a $\ln {\cal C}_{10}$-$\tau_C$-$n_s$
combination; to $\pm 0.02$, $\omega_b$ predominantly, with some extra;
to $\pm 0.04$ a $\ln {\cal C}_{10}$-$\tau_C$-$n_s$-$\omega_{cdm}$
combination, and another to $\pm 0.08$. The sixth, $\pm 0.13$, is in a
direction that would break the amplitude $\tau_C$ degeneracy. When the
$\tau_C$ prior is included, this error diminishes. The least well
determined mode is associated with the $\Omega_k$-$\Omega_\Lambda$
angular-diameter-distance degeneracy.  Goldstein \et\ (2003) give the
eigenmodes for what is basically the \janzerothree\  mix except the \junzerotwo\  CBI
and VSA data are used.  The \janzerothree\  data used here have best determined
modes, to $\pm 0.011$, $\pm 0.015$, $\pm 0.05$, that are mainly similar 
to the \marchzerothree\   ones, but with differing subdominant structure in
the linear combinations.

Whether before or after WMAP, the simplest inflationary paradigm with
minimal parameters fits the data well. This does not mean inflation is
proved, but competitor theories would have to look awfully like
inflation for them to work. As CMB precision increases at high $\ell$,
and more polarization data arrive, and as the LSS and SN1 data
improve, more details with more parameters will be explored. Already
WMAP points once again to the low-$\ell$ anomaly and with the other
data to possible variation in the primordial slope with wavenumber.

{\small\rm We thank our collaborators on the Boomerang, CBI and \Acbar\ teams for
sharing the exhilaration of spectral revelation over the past few
years and Antony Lewis for Cosmomc and other discussions. Lastly we
would have you picture a beaming Dave Wilkinson with Sunyaev and Bond
in front of a Saturn V rocket of the sort that took the US to the
Moon, taken shortly before the Delta-launch of WMAP in June 2001, and the
celebration thereafter --- still ongoing, but greatly missing the W in
WMAP.}

\def\prd{{{\it Phys.~Rev.~D}}}
\def\prl{{{\it Phys.~Rev.~Lett.}}}
\def\apj{{{\it Astrophys.~J.}}}
\def\apjl{{{\it Aatrophys.~J.~Lett.}}}
\def\apjsuppl{{{\it Astrophys.~J.~Supp.}}}
\def\mnras{{{\it Mon.~Not.~R.~Astr.~Soc.}}}
\def\aa{{{\it Astron.~Astrophys.}}}
\def\procroysoc{{{\it Phil.~Trans.~R.~Soc.~Lond.~A}}}

\end{document}